\documentclass[
      aps,
      prb,
      twocolumn,
      superscriptaddress,
      showpacs]{revtex4-1}
     
\usepackage{amsmath}
\usepackage{amsfonts}
\usepackage{amssymb}
\usepackage{amsthm}
\usepackage{graphicx}
\usepackage{bbm}
\usepackage{bm}
\usepackage{bbold}
\usepackage{xcolor}
\usepackage{braket}
\usepackage{hyperref}
\usepackage{soul}
\usepackage{ulem}

\newcommand{\sinc}{{\rm sinc}}

\begin{document}

%----------------------------------------------------------------%
\title{\textcolor{black}{Photoinduced electronic and spin properties of two-dimensional \\ electron gases with Rashba spin-orbit coupling under perpendicular magnetic fields}}

 \author{Daniel Hernang\'{o}mez-P\'{e}rez}
 \affiliation{Institute of Theoretical Physics, University of Regensburg, 93040 Regensburg, Germany}
 
 \affiliation{Department of Materials and Interfaces, Weizmann Institute of Science, Rehovot 7610001, Israel}

\author{Juan Daniel Torres}
 \affiliation{School of Physical Sciences and Nanotechnology, Yachay Tech University, Urcuqui 100119, Ecuador}
\author{Alexander L\'{o}pez}

\email[Corresponding author: ]{alexlop@espol.edu.ec}

 \affiliation{Escuela Superior Polit\'ecnica del Litoral, ESPOL, Departamento de F\'isica, Facultad de Ciencias Naturales y Matem\'aticas, Campus Gustavo Galindo
 Km. 30.5 V\'ia Perimetral, P. O. Box 09-01-5863, Guayaquil, Ecuador}	
\date{\today}

\begin{abstract}
We theoretically investigate photoinduced phenomena induced by time-periodic driving fields in two-dimensional electron gases under perpendicular magnetic fields with Rashba spin-orbit coupling. 
Using perturbation theory, we provide analytical results for the Floquet-Landau energy spectrum appearing due to THz radiation.
By employing the resulting photo-modulated states, we compute the dynamical evolution of the spin polarization function for an initially prepared coherent state.
We find that the interplay of the magnetic field, Rashba spin-orbit interaction and THz radiation can lead to  \textcolor{black}{non-trivial beating patterns in} the spin polarization.
The dynamics also induces fractional revivals in the autocorrelation function due to interference of the photo-modulated quantum states.
\textcolor{black}{We} calculate the transverse photo-assisted conductivity in the linear response regime using Kubo formalism and \textcolor{black}{analyze} the impact of the radiation field and Rashba spin-orbit interaction. In the static limit, we find that our results reduce to well-known expressions of the conductivity in non-relativistic and quasi-relativistic (topological insulator surfaces) two-dimensional electron gas thoroughly described in the literature.
% and the conductivity of topological insulators single surfaces characterized by a single Dirac cone. 
%Finally, 
\textcolor{black}{We} discuss the possible experimental detection of our theoretical prediction and their relevance for spin-orbit physics at high \textcolor{black}{magnetic} fields.
\end{abstract}
\maketitle

%----------------------------------------------------------------%
\section{Introduction}
% Motivation
Over the past years, photoinduced properties of two-dimensional systems have been a subject of tremendous study, especially after the 
identification of non-trivial topological properties that can be induced by periodic driving fields\cite{NP2011Lindner, PRL2018_Refael,PRB2018_Lindner,PRB2017_Refael,PRB2018_Lindner_2,PRB2014_PerezPiskunow_2,1909.02008_Lindner}. 
Another reason for the considerable amount of attention devoted to these systems, both theoretically and experimentally, comes from the fact that driving fields can be used to dynamically control material and topological properties, \textit{i.e.} one can by simply shining monochromatic laser light promote the system to different topological phases (``Floquet engineering'' [\onlinecite{1909.02008_Lindner}]).
In this regard, non-trivial light-induced \textcolor{black}{phenomena} have been shown to
exist in several two-dimensional systems \textcolor{black}{such as standard two-dimensional electron gases\cite{Shelykh2015a, Shelykh2015b, Shelykh2016a, Shelykh2016b},} monolayer graphene\cite{Oka2009,1610Montambaux,1910.13510_Refael,FoaTorres1,PRB2014_PerezPiskunow_2,PRB2014_PerezPiskunow}, silicene\cite{PRL2012Ezawa,NJP2012Ezawa,PRL2013Ezawa},
transition metal dichalcogenides\cite{PRB2019_Burkard}
or topological insulators\cite{Wang2013,Calvo2015, Mahmood2016}.

When the electrons in a two-dimensional plane are additionally subjected to a static perpendicular magnetic field, their periodic motion translate into  discrete (Landau) levels due to quantization of the electronic kinetic energy\cite{Prange1987}. This  quantization \textcolor{black}{lies} at the origin of dramatic consequences for macroscopic transport properties at low temperatures, the most famous being the perfect quantization of the Hall conductance in plateaus of integral multiples of the conductance quantum\cite{Klitzing1980}. 
The robustness of the conductance quantization has been addressed by considering the effect of THz driving\cite{Morimoto2009} or spin-orbit (SO) interaction of the Rashba type\cite{Hernangomez2014, Carbotte2014} among other types of perturbations. The latter appears due to asymmetric confinement of electron gases in low-dimensional nanostructures and can be tuned using local external electric fields\cite{Nitta1997}. A physically relevant and interesting scenario then occurs when both of these tunable pertubations are simultaneously present in the system, which can happen when two-dimensional surface gases existing in materials with heavy atoms such as InSb\cite{Morgenstern2011} or BiSb monolayers\cite{PRB2017Singh} are irradiated by a periodic time-dependent field.
%

% \subsection{Short summary of the results}
In this work, we consider this scenario and study the combined effect of periodic driving (Floquet) and Rashba SO interaction in clean two-dimensional electron gases (2DEG) under high perpendicular magnetic fields. 

Note that, in this situation, one is also required to incorporate the 
effect of the Zeeman coupling, which might affect the structure
of the energy levels.
Thanks to the periodicity of the radiation field, we apply Floquet's theorem and transform
the dynamical equations of motion into an exact time-independent problem. 
Our approach has then the advantage that the dynamics can be tackled without the 
need of addressing an infinite-dimensional eigenvalue problem.
We investigate the emergence of light-modulated Landau energy levels (dubbed Floquet-Landau levels),  similar to the static Landau levels but with radiation renormalizing both the Rashba SO parameter and the Zeeman coupling. 
Using the driven eigenstates, we compute the dynamical evolution of relevant physical observables such as the spin polarization or the autocorrelation function and investigate the effect of SO coupling in the linear response photoconductivity. 
We further use our results for the photoconductivity to explore different physical regimes characterized by the strength of the Rashba SO interaction.
At small values of the Rashba parameter, we recover the results from the ordinary photo-excited 2DEG. At large values of the SO coupling strength, we obtain expressions for the conductivity of graphene / single surface of topological insulators previously described in the literature. Finally, we discuss the possible experimental probe of our theoretical predictions in realistic systems.
%
%

% \subsection{Organization of the paper}
The paper is organized as follows: In Sec. \ref{sec2}, we present the model Hamiltonian and study the effect of the radiation field on the 
spectral properties by using perturbation theory. 
In Sec. \ref{sec3}, we consider relevant observables and study the time-evolution of
the spin polarization and the autocorrelation function when the system is initially prepared in a coherent state. 
In Sec. \ref{sec4}, we obtain the photoconductivity of the Floquet system using Kubo formula and analyze its behavior for several regimes of the effective SO interaction.
Finally, in Sec. \ref{sec:conclusion}  we give concluding remarks. 
We complement the paper by showing explicit algebraic derivations of our main results in the appendix.

%%%%%%%%%%%%%%%%%%%%%%%%%%%%%%%%%%%%%%%%%%%%%%%%%%%%%%%%%%%%%%%%%%%%%%%%%%
\section{Model}\label{sec2}

\subsection{Static Hamiltonian}\label{sec:static}

We consider a single electron of spin $1/2$, electronic charge $q=-e$ (here $e>0$) and effective mass $m^\ast$ 
confined to a two-dimensional (2D) plane under a perpendicular
and uniform magnetic field, ${\bf B}=B\hat{\mathbf{z}}$.
The single-particle Hamiltonian in the presence of SO coupling of the Rashba type and Zeeman interaction is given by
\begin{equation}\label{eq1}
{{H}}_0=\frac{{\bm{\pi}}^2}{2m^\ast} \otimes 1\!\!1_2 +  \lambda [{\bm{\pi}} \times \bm{\sigma}]_z  + \frac{\Delta}{2} \otimes \sigma_z.
\end{equation}
Here, the first term corresponds to the spin-diagonal Hamiltonian for a free single electron, 
with $1\!\!1_2$ being the $2\times2$ identity matrix in spin space, $\bm{\pi}$ the gauge-invariant momenta with components ${\pi}_j=p_j+eA_j(\mathbf{r})$ [$j \in \{x,y\}$, $\mathbf{r} = (x,y)$ is the position of the electron and $\mathbf{A}(\mathbf{r})$ the electromagnetic vector potential]. The later is related to the external magnetic field through the constitutive relation $\nabla_\mathbf{r}\times{\bf A}(\mathbf{r})={\bf B}$. 
The second term is the Rashba Hamiltonian describing the coupling between spin and orbital degrees of freedom 
\begin{equation}
H_\textnormal{R} = \lambda [{\bm{\pi}} \times \bm{\sigma}]_z = \lambda [\pi_x \otimes \sigma_y - \pi_y \otimes \sigma_x], 
\end{equation}
with $\lambda$ being the spatially averaged Rashba parameter and $\bm{\sigma} = (\sigma_x, \sigma_y, \sigma_z)$ the vector of Pauli matrices.
Finally, the third term is the Zeeman coupling between the spin of the electron and the external magnetic field characterized by the Zeeman gap $\Delta$. We omit in what follows the tensor product symbol and the identity matrix $1\!\!1_2$.

Introducing the magnetic length $l_B = \sqrt{\hbar/(e B)}$ and the cyclotron frequency $\omega_c=\hbar/(m^\ast l_B^2)$, as well as the annihilation and creation operators 
\begin{subequations}\label{eq:optransform}
\begin{align}
    a=\frac{l_B}{\sqrt{2}\hbar}(\pi_x-i\pi_y), \\
    a^\dagger=\frac{l_B}{\sqrt{2}\hbar}(\pi_x+i\pi_y),
\end{align}
\end{subequations}
it can be easily shown that the Hamiltonian in Eq. \eqref{eq1} can be rewritten as
the Jaynes-Cummings Hamiltonian\cite{Debald2005} well-known in quantum optics
\begin{eqnarray}\label{eq:JC_Hamiltonian}
{H}_{0}=\hbar\omega_c N_a- \frac{\delta}{2}\sigma_z-i\lambda_B (a\sigma_+-a^\dagger\sigma_-).
\end{eqnarray}
Here $\lambda_B=\sqrt{2}\hbar\lambda/l_B$ characterizes the strength of the SO interaction in the presence of magnetic field (\textit{i.e.} the SO interaction strength per magnetic length), $ \sigma_\pm=(\sigma_x\pm i\sigma_y)/2$, the detuning is given by $\delta= \hbar \omega_c-\Delta$ and we have defined the operator $N_a:=a^\dagger a+(1 + \sigma_z)/2$, which commutes with the Rashba SO Hamiltonian. We assume in this paper that  $\delta >0$ as the cyclotron energy is typically the dominant energy scale in comparison to the Zeeman gap.
After straightforward diagonalization of Eq. \eqref{eq:JC_Hamiltonian}, we get the energy levels (Landau levels\cite{Bychkov1984, Hernangomez2013}) given by  
\begin{equation}\label{energy0}
E_{sn}=\hbar\omega_c n +\frac{s}{2} \Delta_n.
\end{equation}
where $\Delta_n = \sqrt{4n\lambda_B^2+\delta^2}$. 
The energy levels are characterized by a positive integer, $n \geq 0$, the Landau level index
and the SO quantum number $s = s(n)$ which takes the value $s=\pm$ if $n\geq 1$ and $s=+$ when $n=0$. The quantum number $s$ can be interpreted as the ``spin'' index projection along the axis defined by the Rashba SO interaction. 
As in the spinless case, the degeneracy of each level per unit area is equal to $ n_B = 1 /(2\pi l_B^2)$.

\begin{figure}
  	  \centering
  	  \includegraphics[width=0.5\textwidth]{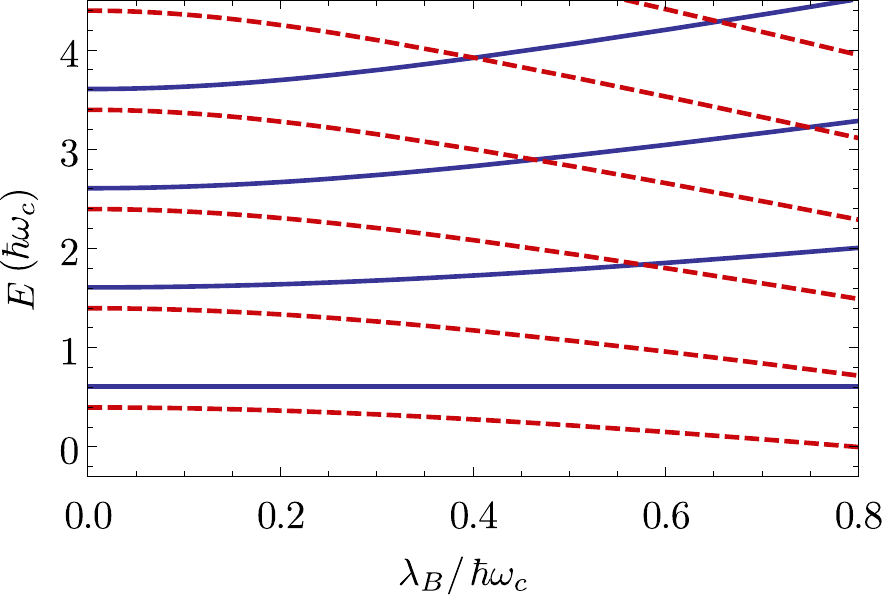}
  	  \caption{Landau levels \eqref{energy0} given in units of the cyclotron energy, $\hbar \omega_c$, as a function of the dimensionless parameter $\lambda_B / (\hbar \omega_c)$. The solid and dashed lines correspond to each of the SO projections labeled by the quantum number $s$. As detailed in the main text, typical parameters for 2DEG on BiSb monolayers
  	  are considered.
  	    }\label{fig:spectrum}
\end{figure}

In Fig. \ref{fig:spectrum}, we show the energy spectrum \eqref{energy0} (normalized to the cyclotron energy) as a function of $\lambda_B/(\hbar \omega_c)$. We have considered values for the electron effective mass, $m^\ast = 0.02 m_0$, normalized detuning, $\delta/(\hbar \omega_c) \simeq 1.2$, and Rashba SO interaction typical for 2DEG that can be found in BiSb monolayers\cite{PRB2017Singh} (which have SO coupling strength $\hbar \lambda = 2.3$ eV$\cdot$\AA).
For small values of the renormalized SO interaction relative to
the cyclotron energy, the energy levels are distributed
as Zeeman-split pairs of Landau levels. Once the effect of the magnetic-field on
the SO coupling becomes relevant, each pair of levels splits off [note that, similar to the case of graphene\cite{Champel2010, Lopez2015} the level with quantum numbers $(0,+)$ is independent of the SO coupling].
The energy spectrum then presents non-equidistant levels. Consequently, it also exhibits level crossings between different pairs
of Landau levels $(n,s)$ and $(n', -s)$ [\onlinecite{Hernangomez2013}]. Note that necessarily energy levels with the same SO quantum number never cross and that there is a unique energy level labeled by the quantum numbers $(1,-)$ \textcolor{black}{that} never crosses with any other Landau level. The accidental degeneracies produced by the level crossings are expected to be lifted once Landau level mixing occurs due to the disorder potential.

% 

%%%
% 
The eigenstates of the spinful static Hamiltonian can be written in terms of the eigenstates $|n\rangle$ of the operator $a^\dagger a$ (\textit{i.e.} eigenstates of the spinless Hamiltonian with the vector potential expressed in the Landau gauge, $\mathbf{A} = -B y \hat{\mathbf{x}}$). We find
\begin{eqnarray}\label{ev0}
|\phi_{sn}\rangle=\left(
\begin{array}{c}
-is c_{-sn}|n-1 \rangle \\
 c_{sn}|n \rangle
\end{array}
\right),
\end{eqnarray}
 where
\begin{equation}\label{cs}
c_{sn}=\sqrt{\frac{\Delta_n+s\delta}{2\Delta_n}}.
\end{equation}
Here, we set $|-1 \rangle \equiv 0$ so that Eq. \eqref{ev0} also holds
for the lowest Landau level. 
To simplify the notation, we have noted $|n\rangle \equiv |n,k \rangle$ where
$k \equiv k_x$ with $0 < k < (2\pi / L_x) n_B$  is a continuous quantum number characterizing the degeneracy of each Landau level.
Due to translational invariance of the observables under consideration in the next sections $k$ will always remain a good quantum number and we therefore omit any reference to it in what follows. 

\subsection{Effect of \textcolor{black}{electromagnetic} radiation} 
Let us now consider the effect of circularly polarized electromagnetic radiation, incident perpendicularly to the 
sample. 
We assume that the beam radiation spot is large enough compared to the lattice spacing so we can neglect 
any spatial variation of the incident beam. As such, the resulting light-matter interaction can be described by means of a 
homogeneous time-dependent vector potential
\begin{equation}
\bm{\mathcal{A}}(t)=\frac{\mathcal{E}}{\Omega}\left(\cos\Omega t,\sin\Omega t\right),
\end{equation}
where $\mathcal{E}$ and $\Omega$ are, respectively, the amplitude and frequency of the electric field. The expression of the incident field can be easily obtained from the standard relation
$\bm{\mathcal{E}}(t)=-\partial_t \bm{\mathcal{A}}(t)$. 
For simplicity, we also assume that the light beam is right-handed circularly polarized, extension to left-handed circular polarization
being 
\textcolor{black}{straightforward by considering the transformation $\Omega\rightarrow -\Omega$}. 
\textcolor{black}{Linearly polarized driving electromagnetic radiation  can also be considered in the present scheme. Compared to the circular polarization, the 
linearly polarized electromagnetic field carries no orbital angular momentum but breaks the rotational invariance of the 2DEG. In graphene without external magnetic fields\cite{Wu2011, Scholz2013}, circular polarization opens additional gaps compared to the linear case. Because here the spectrum is already gapped due to the magnetic field, the distinction between both polarization states becomes less relevant and we proceed our analysis by considering circularly polarized radiation only.}

Thus, considering the total vector potential $\bm{\mathcal{A}}(\mathbf{r}, t)=\bm{A}(\mathbf{r})+\bm{\mathcal{A}}(t)$, we apply the minimal coupling prescription $\mathbf{p}\rightarrow \mathbf{p}+ e{\bm{\mathcal{A}}}(\mathbf{r}, t)$ in order to obtain an interaction potential term with the driving field
\begin{equation}
V(t)=\xi( \sigma_y \cos \Omega t-\sigma_x \sin\Omega t). 
\end{equation}
This term enters into the time-dependent Hamiltonian ${H}(t)={H}_0+V(t)$
with an effective coupling constant $\xi=e\lambda\mathcal{E}/\Omega$. 
Observe that within our approach the additional time-dependent contribution
independent of the SO coupling term is neglected. It can be easily checked that this is valid under the assumption $\omega_c < \Omega$, which holds in part of the THz and infrared spectral range for not very large magnetic field strength. 
Inclusion of the $V(t)$ term makes the full Hamiltonian $H(t)=H_0+V(t)$
periodic in time, $H(t+T)=H(t)$, with $T=2\pi/\Omega$ being the period of oscillation of the driving field. 
We now apply Floquet's theorem and write the evolution operator of the system in the form\cite{Grifoni2009} 
\begin{equation}\label{unifull}
U(t)=P(t)e^{-iH_F t}, 
\end{equation}
with $P(t)$ a periodic unitary matrix and $H_F$ a time-independent dynamical generator referred to as the Floquet Hamiltonian. The eigenvalues of the Floquet Hamiltonian $H_{F}$ form the so-called {quasienergy} spectrum  of 
the periodically driven system. 

For the system under consideration, it can be shown that $P(t)=\exp(-iN_a\Omega t)$ generates a time-dependent unitary transformation, 
$|\Psi(t)\rangle=P(t)|\Phi(t)\rangle$, such that the time-dependent Schr\"{o}dinger equation 
\begin{equation}\label{evolution}
i\hbar\partial_t|\Psi(t)\rangle=H(t)|\Psi(t)\rangle,
\end{equation}
becomes
\begin{equation}\label{effec1}
i\hbar\partial_t|\Phi(t)\rangle=H_{F}|\Phi(t)\rangle,
\end{equation}
where $|\Phi(t)\rangle$ are the Floquet eigenstates.
Doing the explicit calculation (see Appendix \ref{app:perturbative_terms}),
$H_F$ is found to be given by
\begin{equation}\label{heff}
H_{F}=\hbar  \omega_{-} N_a -\frac{\delta}{2}\sigma_z+i\lambda_B (a^\dagger\sigma_--a\sigma_+)
-\xi\sigma_y,
\end{equation}
where we have introduced the frequency $\omega_{-} = \omega_c - \Omega$.
We first notice that when the resonant condition, $\Omega=\omega_c$, is fulfilled the resonant Hamiltonian $H_{r}$
expressed in terms of shifted operator $b=a-\beta$, with $\beta=\xi/\lambda_B$, can be written as %
\begin{equation}\label{hr}
H_{r}=i\lambda_B (b^\dagger\sigma_--b\sigma_+)-\frac{\delta}{2}\sigma_z.
\end{equation}
Therefore, when the resonance condition is satisfied, the spectrum is the same as in Eq. \eqref{energy0}, but an integer number of excitations $\hbar \omega_{-}$ have been resonantly absorbed from the system. This would be an $m$ photon resonance.

When the system is not at resonance, we apply perturbation theory to obtain an effective Floquet Hamiltonian that allows us to study the full frequency response of the system.
We consider as small parameter the effective radiation strength $\kappa=\xi/\hbar\omega_c$ and transform the Hamiltonian
in Eq. \eqref{heff} as $H=\exp[-i(\kappa/2) I_+] H_F \exp[i(\kappa/2) I_+]$
where the operator $I_+=a^\dagger\sigma_-+a\sigma_+$ commutes with ${N}_a$.
Evaluation up to first order in $\kappa$ using the Baker-Campbell-Hausdorff formula (see Appendix \ref{app:perturbative_terms}) gives
%, at leading order in $\kappa$,
the effective Floquet
Hamiltonian
\begin{align}
H^\textnormal{eff}_F& \simeq \hbar\omega_- N_c-\Big( \frac{\delta-2\kappa\lambda_B N_c}{2}\Big)\sigma_z  \notag \\
& +i\Big(\lambda_B+\frac{\kappa\delta}{2}\Big) (c^\dagger\sigma_--c\sigma_+),\label{hfin}
\end{align}
where $c=a-\gamma$, $\gamma=2\xi/(\kappa\delta+2\lambda_B)$ and the shifted number operator $N_c= c^\dagger c + (1 + \sigma_z)/2$. Higher order terms in $\gamma$ and $\kappa$ can be dealt, in principle, by using higher order perturbation theory.
The condition $\kappa=\xi/\hbar\omega_c\ll1$ can be easily met for realistic systems.
To check this explicitly, let us take as typical values for the parameters 
 $\mathcal{E} \simeq 0.15$ MV/m, $\hbar\Omega\simeq 10-20$ meV [\onlinecite{PRLKarch2010}], and Rashba coupling constant %$ \hbar \lambda = 2.3\,\textrm{eV}\, \text{\AA}$ 
typical for 2DEG existing on BiSb monolayers\cite{PRB2017Singh} 
 or InSb surface gases ($ \hbar \lambda = 0.7\,\textrm{eV}\, \text{\AA}$) \cite{Morgenstern2011}. This yields for a magnetic field of $B=1$ T values for the perturbative parameter $\kappa\simeq 10^{-2}-10^{-3}$.

 % 5.25 meV BiSb for 1T cyclotron energy effective mass m = 0.0187 m_e
 % 3 meV InSb for 1T cyclotron energy, effective mass m = 0.035 m_e
 % typical SOC energy is E = m \lambda^2
 
\subsection{Floquet-Landau energy spectrum}
We proceed by diagonalization of the effective Hamiltonian \eqref{hfin} in order to obtain the discrete Floquet-Landau energy spectrum (\textit{i.e.} Landau levels dressed by the radiation)
\begin{equation}\label{quasi}
\varepsilon_{sm}= m \hbar\omega_{-} + \frac{s}{2}\sqrt{4 m \tilde{\lambda}^2_B+\tilde{\Delta}_m^2},
%\,~\mod\hbar\omega_{-},
\end{equation}
with $\tilde{\lambda}_B =\lambda_B + {\kappa\delta}/2$ and $\tilde{\Delta}_m= \delta - 2m\kappa\lambda_B$.
As in the static case, Sec. \ref{sec:static}, the (Floquet) band index $m \geq 0$ is an integer (that plays the role of the Landau level index) and
the new SO quantum number $s=s(m)$ is equal
to $s=\pm$ if $m \neq 0$ and $s=+$ when $m=0$.
By comparing this result to Eq. \eqref{energy0}, we find that the radiation field 
affects both the strength of the SO and the (Zeeman-related) detuning: the magnetic field renormalized SO interaction increases due to coupling to the Zeeman term via the radiation; correspondingly, the detuning is reduced by the SO coupling and becomes dependent on the Floquet band index. 

\begin{figure}
  	  \centering
  	  \includegraphics[width=0.5\textwidth]{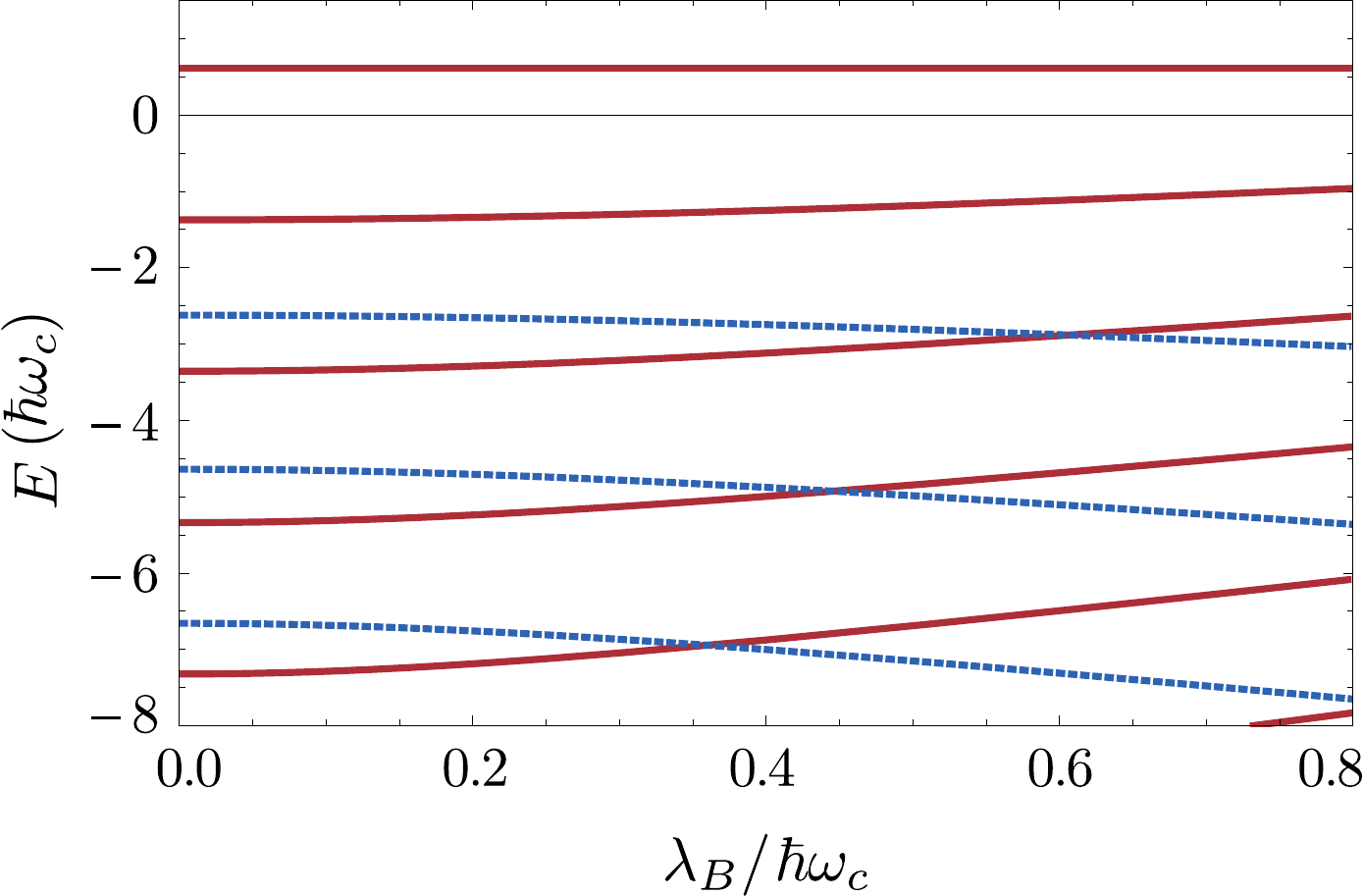}
  	  \caption{Floquet-Landau spectrum \eqref{quasi} given in units of the cyclotron energy, $\hbar \omega_c$, as function of the dimensionless parameter $\lambda_B/(\hbar \omega_c)$. Similarly to Fig. \ref{fig:spectrum}, the solid and dotted lines correspond to the two projections labeled by the quantum number $s$. We consider values of the parameters typical for BiSb 2DEGs (Ref. \onlinecite{PRB2017Singh}), \textcolor{black}{$\kappa = 0.25$ and $\Omega = 3 \omega_c$.}}\label{fig:spectrum_floquet}
\end{figure}

We show in Fig. \ref{fig:spectrum_floquet} the Floquet-Landau spectrum
for the same set of parameters used in Fig. \ref{fig:spectrum}.
Similar to the static Landau energy levels, the Floquet-Landau
energies present multiple level crossings between pairs of levels $(m,s)$ and $(m',-s)$ occurring as a function of 
the SO coupling $\lambda$ or the magnetic field.
Due to the radiation field, the position of the crossings is 
drastically altered.
In addition, most of the energy levels are substantially shifted in energy by the
radiation except
$(0,+)$ which remains unaffected.
Note also that, as a consequence of the shift in the Floquet-Landau levels, the
 levels with positive SO projection $(0, +)$ and $(1, +)$ never
cross with any other energy level.

The corresponding Floquet eigenstates for any Floquet-Landau level labeled by $(m,s)$ can be written in a similar way to Eq. \eqref{ev0} and read
\begin{equation}\label{evfull}
|\psi_{sm}\rangle=\left(
\begin{array}{c}
-is b_{-sm}|m-1\rangle\\
  b_{sm}|m\rangle
\end{array}
\right),
\end{equation}
where we have defined the coefficients
\begin{equation}\label{bs}
b_{sm}=\sqrt{\dfrac{1}{2} \left(1 + s \dfrac{\tilde{\Delta}_m}{\varepsilon_m}\right)},
\end{equation}
with $\varepsilon_{m} =|\varepsilon_{sm} - m\hbar \omega_{-}| $. Observe that the energy level degeneracy given by the continuous quantum number $k$ is not affected and therefore it will remain implicit in our expressions.

\section{Spin and autocorrelation function dynamics}\label{sec3}
\subsection{Spin polarization}
We begin by considering the dynamics of the spin-polarization, whose time-average $\langle\sigma_z\rangle$ can be defined by
\begin{equation}\label{sigmaz}
\langle\sigma_z\rangle= \dfrac{1}{T}\int^T_0dt\bra{\Psi(0)}U^\dagger(t)\sigma_zU(t)\ket{\Psi(0)}, 
\end{equation}
where $\ket{\Psi(0)}$ is the state of the system prepared at $t=0$. Taking
into account that $[\sigma_z,P(t)]=0$, we use Eq. \eqref{unifull} 
to further write the last expression using the Floquet Hamiltonian
\begin{equation}
\langle\sigma_z\rangle=\dfrac{1}{T} \int^T_0dt\bra{\Psi(0)}e^{iH_F t/\hbar}\sigma_ze^{-iH_F t/\hbar}\ket{\Psi(0)}. 
\end{equation}
Note that, whenever the initial state is an eigenstate of the Floquet Hamiltonian, the expectation value of the spin polarization in Eq. \eqref{sigmaz} is constant over one period of the radiation field. %

\textcolor{black}{
Yet, with an experimental setup in mind, it is more feasible to prepare the initial state of the system in a linear combination of
eigenstates of the static Hamiltonian \eqref{eq1}. Several initial states are possible (for instance, just Eq. \eqref{ev0}, superposition of thermally occupied Landau levels, \textit{etc.})
but here we consider the experimentally relevant case at low temperatures of $\ket{\Psi(0)}$ being a coherent state superposition of Landau levels.
For the coherent state, $\ket{\alpha}$, the coherent state parameter $\alpha = \sqrt{\braket{\alpha | N_a| \alpha}}$ gives 
the mean number of Landau levels that are excited, $\alpha$ being in this context analogous 
to the mean photon number in quantum optics\cite{agarwal_2012}.}
The coherent state is explicitly given by the expression
\begin{equation}\label{coherent}
\ket{\alpha}=e^{-\frac{\lvert\alpha\rvert^2}{2}}\Big(\ket{\varphi_{0}}+\frac{1}{\sqrt{2}}\sum_{n = 1}^{+\infty} \sum_s \frac{\alpha^n}{\sqrt{n!}}\ket{\varphi_{sn}}\Big),
\end{equation}
with $\ket{\varphi_{sn}}$ being the eigenstates of the static Hamiltonian at zero detuning ($\delta=0$),
\begin{equation}\label{evzero}
|\varphi_{sn}\rangle=\frac{1}{\sqrt{2}}\left(
\begin{array}{c}
-is|n-1\rangle\\
  |n\rangle
\end{array}
\right),
\end{equation}
for $n\neq 0$ whereas $\ket{\varphi_{+0}}=\ket{\phi_{+0}}$. 

\begin{figure}
  	  \centering
  	  \includegraphics[width=0.5\textwidth]{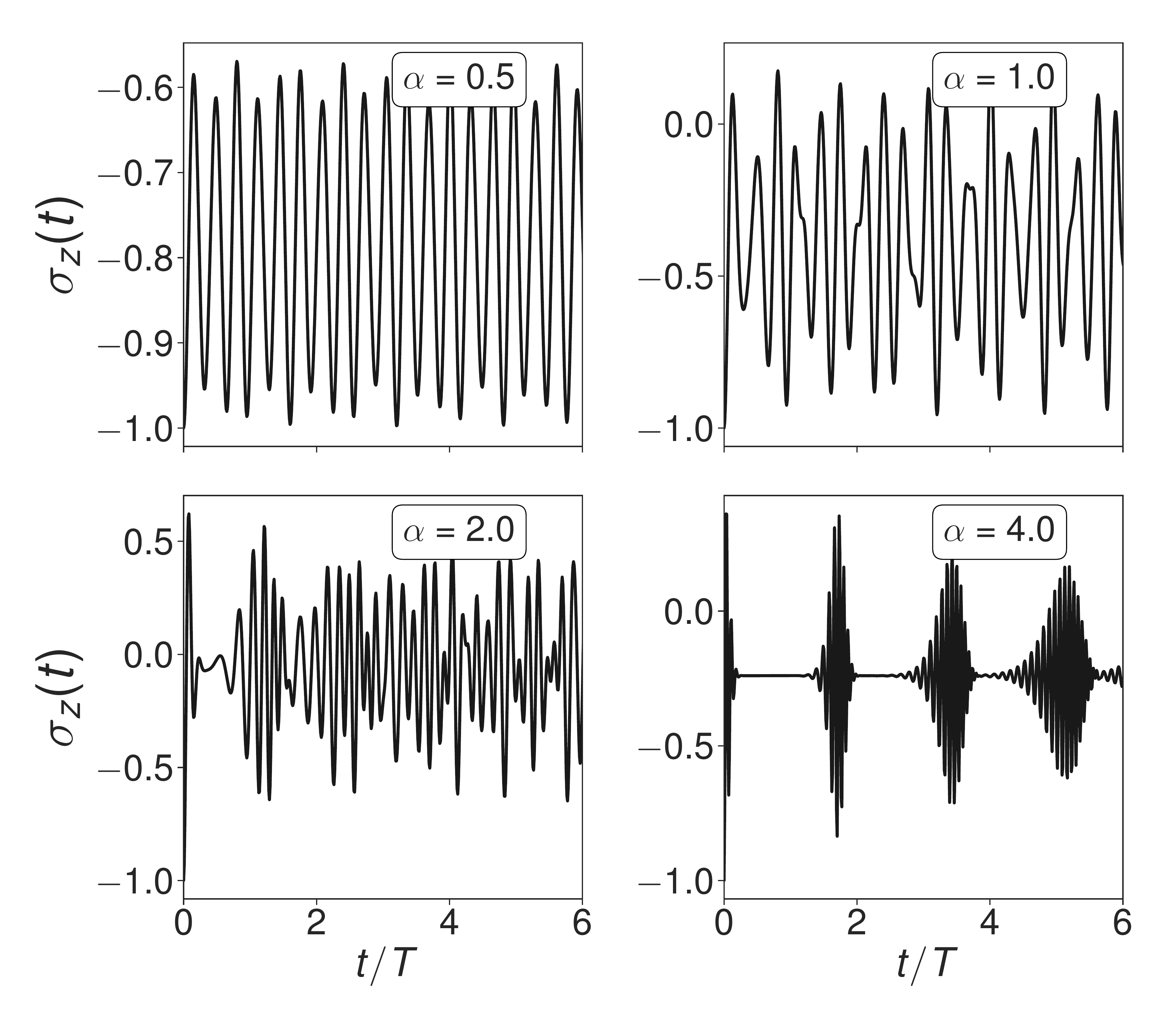}
  	  \caption{Time evolution of the spin polarization, $\sigma_z(t)$, obtained for different values of the coherent state parameter $\alpha$. We consider the effective radiation strength $\kappa=0.25$ and set $\lambda_B/\hbar\Omega=\delta/\hbar\Omega=1$ and $B=1$ T.
  	    }\label{fig:f00}
\end{figure}
 
Observe that the coherent state also satisfies $\braket{\alpha | \sigma_z|  \alpha}=-1$. Thus, any change (oscillation, decay, \textcolor{black}{ etc.}) in the spin polarization as a function of time is either related to spin flipping due to Rashba SO interaction or to fluctuations in the excitation number due to the periodic driving.
This initial state has been considered previously to study
photoinduced effects on the Landau levels of monolayer
graphene.\cite{Lopez2015}
We now explore the effect on the dynamics of ``real'' spin by the interplay of photoinduced renormalized Rashba SO interaction and Zeeman term.

After straightforward calculations  (see Appendix \ref{app:details_spinpolarization} for detailed derivations), the time-dependent spin polarization $\sigma_z(t) = \bra{\alpha}U^\dagger(t)\sigma_z U(t)\ket{\alpha}$ for the coherent state is given by the expression
\begin{align}\label{eq:spin_polarization}
    \sigma_z(t) &=-e^{-\lvert\alpha\rvert^2}\Bigg(1+\sum_{m=1}^{+\infty}\frac{\lvert\alpha\rvert^{2m}}{m!} \Bigg\{\Big(\frac{\tilde{\Delta}_m}{\varepsilon_m}\Big)^2+ \notag \\
    & + \Bigg[1-\Big(\frac{\tilde{\Delta}_m}{\varepsilon_m}\Big)^2\Bigg]\cos\left(\dfrac{2 \varepsilon_m t}{\hbar}\right)\Bigg\}\Bigg).
\end{align}

We plot Eq. \eqref{eq:spin_polarization} in Fig. \ref{fig:f00} 
considering an effective dimensionless radiation strength $\kappa=0.25$, along with $\lambda_B/\hbar\Omega=\delta/\hbar\Omega=1$ and $B=1$ T, for different values of the coherent state parameter $\alpha$. %$\alpha=\sqrt{\bra{\alpha}N_a\ket{\alpha}}$. 
At low values of $\alpha$ [panels (a) and (b) in Fig. \ref{fig:f00}], we find small amplitude Rabi oscillations.
This is due to the fact that the dynamics is  mostly dictated by the interference of the lowest Floquet-Landau levels. However, for larger values of $\alpha$, the higher Floquet-Landau levels contribute with larger weight to the interference and dynamical localization effects appear. This result is explicitly shown in Fig. \ref{fig:f00} (c) and (d) where a strong beating pattern as a function of time is present. This situation is qualitatively similar
to the behavior of the pseudo-spin polarization in graphene under periodic illumination\cite{Lopez2015}. However, here the dominant time scale for the 
dynamical localization is not related to the cyclotron frequency but to the photon frequency, $T= 2\pi /\Omega$.
It is also interesting to observe that when collective behavior for the charge carriers sets in the 
spin polarization can change its sign from $\sigma_z(0) = -1$ (\textit{i.e.} spin down) to $\sigma_z(t) > 0$.
\textcolor{black}{In other words, induced by the periodic driving, the exchange of angular momentum between orbital and spin
degrees of freedom mediated by the SO interaction can temporarily inverse the sign of the spin density.}

\begin{figure}
  	  \centering
  	  \includegraphics[width=0.5\textwidth]{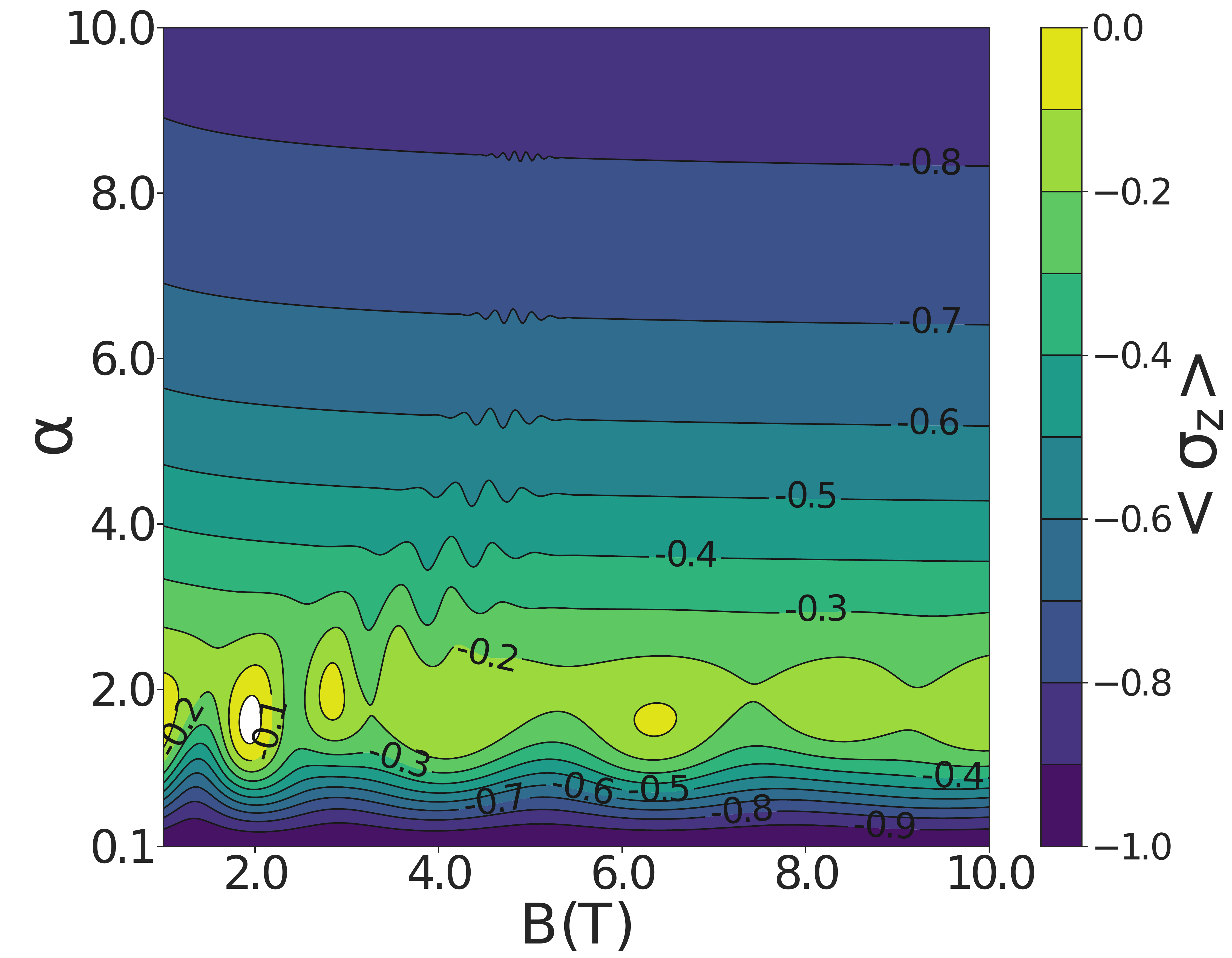}
  	  \caption{Expectation value of the spin polarization, $\langle \sigma_z \rangle$ in $B-\alpha$ parameter space. We consider an effective radiation strength $\kappa=0.25$ and fix the radiation frequency of the incident light beam by setting $m^{*}\lambda^2/\hbar \Omega = 1$.}\label{fig:f0}
\end{figure}

\begin{figure}
  	  \centering
  	  \includegraphics[width=0.5\textwidth]{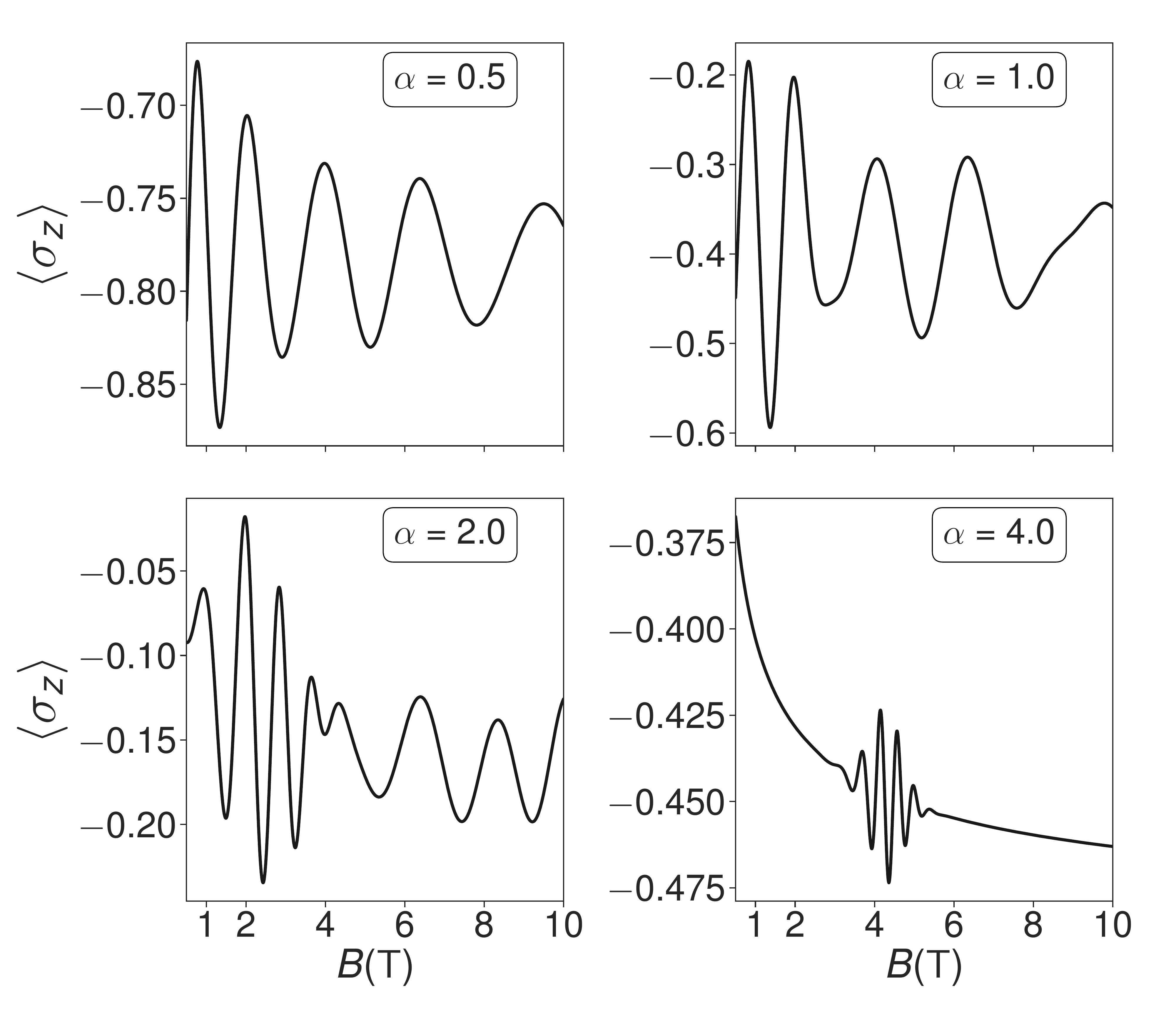}
  	  \caption{Magnetic field response of the expectation value of the spin polarization at four representative values of the coherent state parameter $\alpha$. Values of the effective radiation strength, $\kappa$, and the radiation frequency, $\Omega$, are set as in Fig. \ref{fig:f0}.}\label{fig:f01}
\end{figure}

We can get further insight on the interplay of the  Rashba SO interaction and the radiation field by calculating the mean polarization $\langle\sigma_z\rangle$. This quantity is obtained by averaging the expression (\ref{eq:spin_polarization}) over one period of oscillation of the radiation field [see Eq. \eqref{sigmaz}] and reads 

\begin{align}\label{eq:spin_polarization2}
    \langle \sigma_z\rangle &=-e^{-\lvert\alpha\rvert^2}\Bigg(1+\sum_{m=1}^{+\infty}\frac{\lvert\alpha\rvert^{2m}}{m!} \Bigg\{\Big(\frac{\tilde{\Delta}_m}{\varepsilon_m}\Big)^2 \notag \\
    &  + \frac{2\pi}{\Omega}\Bigg[1-\Big(\frac{\tilde{\Delta}_m}{\varepsilon_m}\Big)^2\Bigg]\sinc\Big(\frac{4\pi\varepsilon_m}{\hbar\Omega}\Big)\Bigg\}\Bigg).
\end{align}
In Fig. \ref{fig:f0} we show a contour density plot of $\langle \sigma_z \rangle$ in the $B-\alpha$ parameter space. As in Fig. \ref{fig:f00}, we set $\delta/\hbar\Omega=1$, $\kappa=0.25$ and consider the photon energy such that $m^{*}\lambda^2/\hbar \Omega = 1$. For BiSb surface gases, this corresponds to THz radiation where $\hbar \Omega \simeq 10$ meV. The general trend observed is that at any value of the static magnetic field $B$, a large magnitude of the average polarization $\langle\sigma_z\rangle$ is achieved for both small and large values of the mean occupation (coherent state parameter) $\alpha$. The value of $\langle \sigma_z \rangle$ is always negative, \textit{i.e.} there is no polarization inversion in the average signal. 
For \textcolor{black}{small and} intermediate values of $\alpha$ the contribution of more Floquet-Landau levels
yields clear oscillations of $\langle \sigma_z \rangle$ as a function of $B$. 
We illustrate this behavior of $\langle \sigma_z \rangle$ by taking ``snapshots'' at fixed values of the mean occupation parameter in Fig. \ref{fig:f01}.

%%%%%%%%%%%%%%%%%%%%%%%%%%%%%%%%%%%%%%%%%%%%%%%%%%%%%%%%%%%%%%%%%%%%%%%%%%%%%
\subsection{Autocorrelation function}

\begin{figure}
  	  \centering
  	  \includegraphics[width=0.5\textwidth]{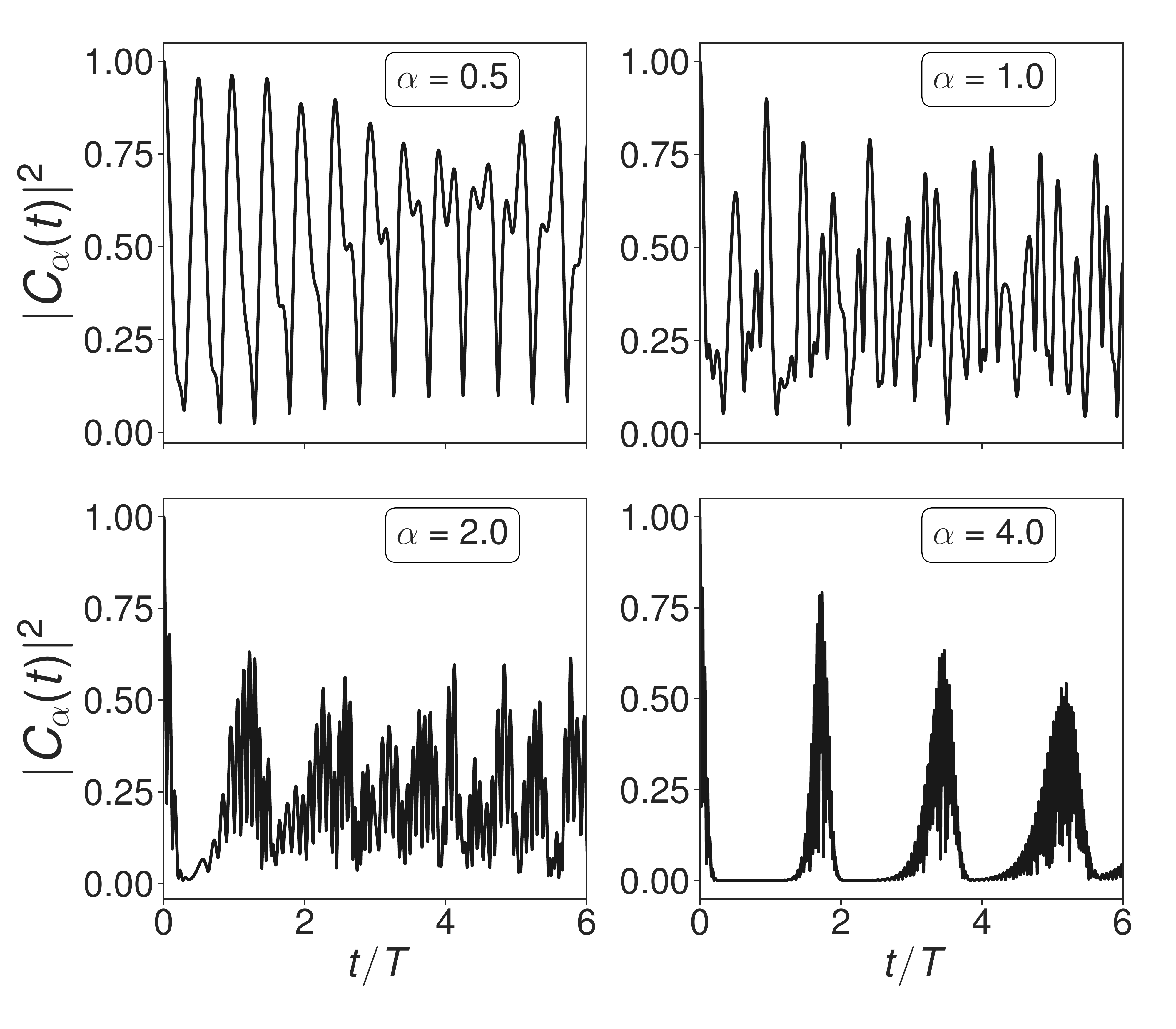}
  	  \caption{Time dependence of the autocorrelation function $C(t)$ as given in Eq. \eqref{eq:autocorrelation} for different values of the coherent state parameter $\alpha$. For larger values of $\alpha$, the autocorrelation function shows partial revivals correlated with the oscillations in the spin polarization seen in Fig. \ref{fig:f00}. Values of the effective radiation strength, $\kappa$, the radiation frequency, $\Omega$, and the magnetic field are set as in Fig. \ref{fig:f00}.}\label{fig:auto}
\end{figure}

We complement the physical picture of the spin dynamics by looking
at the autocorrelation function $ C(t)= \braket{\Psi(0)|\Psi(t)}=\bra{\Psi(0)}U(t)\ket{\Psi(0)}$,
which is simply the overlap between the initial and the time-evolved wave packet.
The absolute value of $C(t)$ provides additional insight on (fractional) quantum revivals induced by the dynamics whenever the time-dependent overlap is close to its maximum value\cite{Romera2009, Romera2011}.
Analogously to the case of the spin polarization, we consider the coherent state \eqref{coherent} and compute $C(t) = \braket{\alpha | \alpha(t)}$. 
For that purpose, we first compute the time-evolved coherent state, $\ket{\alpha(t)}$,
by considering the projection of the dynamics into the basis of the static Hamiltonian. The time evolution of the coherent state is then expressed as
\begin{equation}
 \ket{\Psi(t)} = \sum_{s'} f_m^{ss'}(t) {\rm e}^{-i \varepsilon_m t/ \hbar} \ket{\psi_{s'm}},
\end{equation}
with
\begin{equation}
 f_m^{s s'} = c_{-sn} b_{-s'n} + s s' c_{sn}b_{s'n}.
\end{equation}
After straightforward algebra (see details in Appendix \ref{app:detains_autocorrelation}) we find that the coherent state autocorrelation function
adopts the form
\begin{align}
 C(t) &= {\rm e}^{-|\alpha|^2} \Bigg(1 + \sum_{m=1}^{+\infty} \dfrac{|\alpha|^{2m}}{m!} 
            \Big\{\cos \left(\dfrac{\varepsilon_m t}{\hbar} \right)  \notag \\
            &- i \left[\dfrac{\delta}{\Delta_m} \sqrt{1 - \left(\dfrac{\tilde{\Delta}_m}{\varepsilon_m}\right)^2} +
            \dfrac{\tilde{\Delta}_m}{\varepsilon_m} \sqrt{1 - \left(\dfrac{\delta}{\Delta_m}\right)^2}\right] \notag \\
            &\times \sin \left(\dfrac{\varepsilon_m t}{\hbar} \right) \Big \}    
            \Bigg).\label{eq:autocorrelation}
\end{align}
It can be easily checked that this expression reduces to the result quoted in Ref. \onlinecite{Lopez2015} for graphene in the limit $\lambda_B / (\hbar \omega_c) \gg 1$.

We plot the time evolution of $|C_\alpha(t)|^2$ in Fig. \ref{fig:auto}. To make the comparison with the time-dependent spin polarization explicit, we consider the same values of $\alpha$ as in Fig. \ref{fig:f00}, as well as the same parameters. 
For small values of $\alpha$, as shown in Fig. \ref{fig:auto} panels (a) and (b), the autocorrelation
function presents a strong oscillations reminiscent of the Rabi oscillations
between the two lowest Floquet bands.
For the considered radiation frequency, our result is different from graphene\cite{Lopez2015} since
here the Rabi oscillations involve the quantum interference of more than two Floquet-Landau levels as it can be seen in the beating pattern of $|C_\alpha(t)|^2$.
For larger values of $\alpha$, see Fig. \ref{fig:auto} panels (c) and (d), and
especially for $\alpha \geq 4.0$, the autocorrelation function shows clear fractional
revivals. These revivals are correlated in time with the dynamical localization of the spin polarization shown in Fig. \ref{fig:f00}.
The fractional revivals occur periodically with a period roughly equal to $T_R \simeq 7 T/4$ when the mean Landau level occupation  increases. This means that a full reconstruction of the wavepacket never occurs due to the presence of dephasing.
In the present case, compared to a purely relativistic energy spectrum, this dephasing is produced to the radiation-dressed detuning $\Delta_m$. 
After several periods, the dephasing between the components of the wave packet increases and the value of $|C_\alpha(t = pT_R)|^2$ with $p \in \mathbb{N}_{>0}$ decreases  almost in a linear fashion. 
This decrease in the amplitude of the local maxima of the autocorrelation function in time is a manifestation of the broadening of the spin polarization signal observed in Fig. \ref{fig:f00} (d) for times $t/T_R \geq 2$.

%%%%%%%%%%%%%%%%%%%%%%%%%%%%%%%%%%%%%%%%%%%%%%%%%%%%%%%%%%%%%%%%%%%%%%%%%%%%%%%%%%%%%%%%%%%%%%%%%%%

\section{Photoconductivity}\label{sec4}

We turn now our attention to the transverse photoconductivity, $\sigma_{xy}(\Omega)$, 
computed using the Kubo formula (see details for the derivation in Appendix \ref{app:details_conductivity}).
This approach has already been  employed in the study of the quantum oscillations produced by microwave-induced zero resistance states \cite{Kunold2005}.
Using the Floquet-Landau
basis \eqref{evfull}, we find
\begin{align}\label{eq:HallC}
 	&\sigma_{xy}(\Omega) = \frac{e^2}{h} (\hbar \omega_c)^2 \sum_{m=0}^{+\infty} \sum_{ss'} \dfrac{n_\textnormal{{F}} (\varepsilon_{s'm+1}) - n_\textnormal{{F}} (\varepsilon_{sm})}{(\varepsilon_{s'm+1}- \varepsilon_{sm})^2 - (\hbar \Omega + i \Gamma)^2} \notag
 	\\ &\times
 	\Big[  \sqrt{m} B^{ss'}_{mm+1} +ss'\sqrt{m+1} B^{-s-s'}_{mm+1} 
  	 + \frac{\lambda_B}{\hbar \omega_c} s' B^{s-s'}_{mm+1} \Big]^2
 \end{align}
where $\Omega$ is the photon frequency and $\Gamma$ is a parameter describing the effective Floquet-Landau level broadening due to residual scattering of the electron with impurities\cite{Carbotte2013, Carbotte2014, Mitra2015, Zubair2018}.
To simplify the notation, we have defined in Eq. \eqref{eq:HallC}
 the combination of wavefunction weights $B^{ss'}_{mm'} = b_{sm}b_{s'm'}$, and denoted by 
 \begin{equation}\label{eq:FD}
 n_\textnormal{F}(E)=\dfrac{1}{1+\exp[(E-\mu)/(k_B T)]},
 \end{equation}
 the Fermi-Dirac distribution at temperature $T$ and constant chemical potential $\mu$ ($k_B$ denotes the Boltzmann constant).

Two important observations are now in order. First, employing a (quasi)-equilibrium Fermi-Dirac distribution instead of the non-equilibrium distribution expected\cite{Oka2015, Balseiro2018} due to the driving EM field in Eq. \eqref{eq:HallC} is a good approximation only if (i) the Floquet energies are well-separated in the energy space (ii) the driving frequencies are off-resonance with the transition frequencies and (iii) the laser amplitudes are small.\cite{Mitra2015, Oka2015} Under these circumstances, the driving leads to a renormalization of the parameters of the (time-independent Floquet) system but does not change the distribution function. %, which is evaluated for the Floquet-Landau eigenenergies. 
 In our case (i) holds because the spectrum is always gapped far from the accidental degeneracies, (ii) is achieved by choosing appropriately $\Omega$ to be incommensurate with the transition frequency between adjacent Floquet-Landau levels and (iii) is satisfied from our perturbative approach. Note that incommensurability can be achieved easily as the Floquet-Landau are not equidistant and the separation between them can be tuned with the SO interaction.
Moreover, in 2DEG under strong perpendicular magnetic fields and subjected to time-dependent periodic radiation electron-phonon coupling becomes subdominant at low temperatures. It has also been shown that
the solution of the Boltzmann equation yields at first order a Fermi-Dirac distribution evaluated in the Floquet-Landau levels\cite{Kunold2005}, consistently with the conditions (i)-(iii) discussed above.
Second, the assumption of the broadening of the Floquet-Landau levels to be constant can be justified at high magnetic fields for which the disorder becomes smooth on the scale of the magnetic length. For low magnetic fields and low temperatures, in 
the regime of Shubnikov-de Haas oscillations, it has been shown that both the magnetic field and radiation renormalize the scattering rate and can produce quantitative (but not qualitative) changes of the density of states \cite{Shelykh2016a}.

\begin{figure}
  	  \centering
  	  \includegraphics[width=0.45\textwidth]{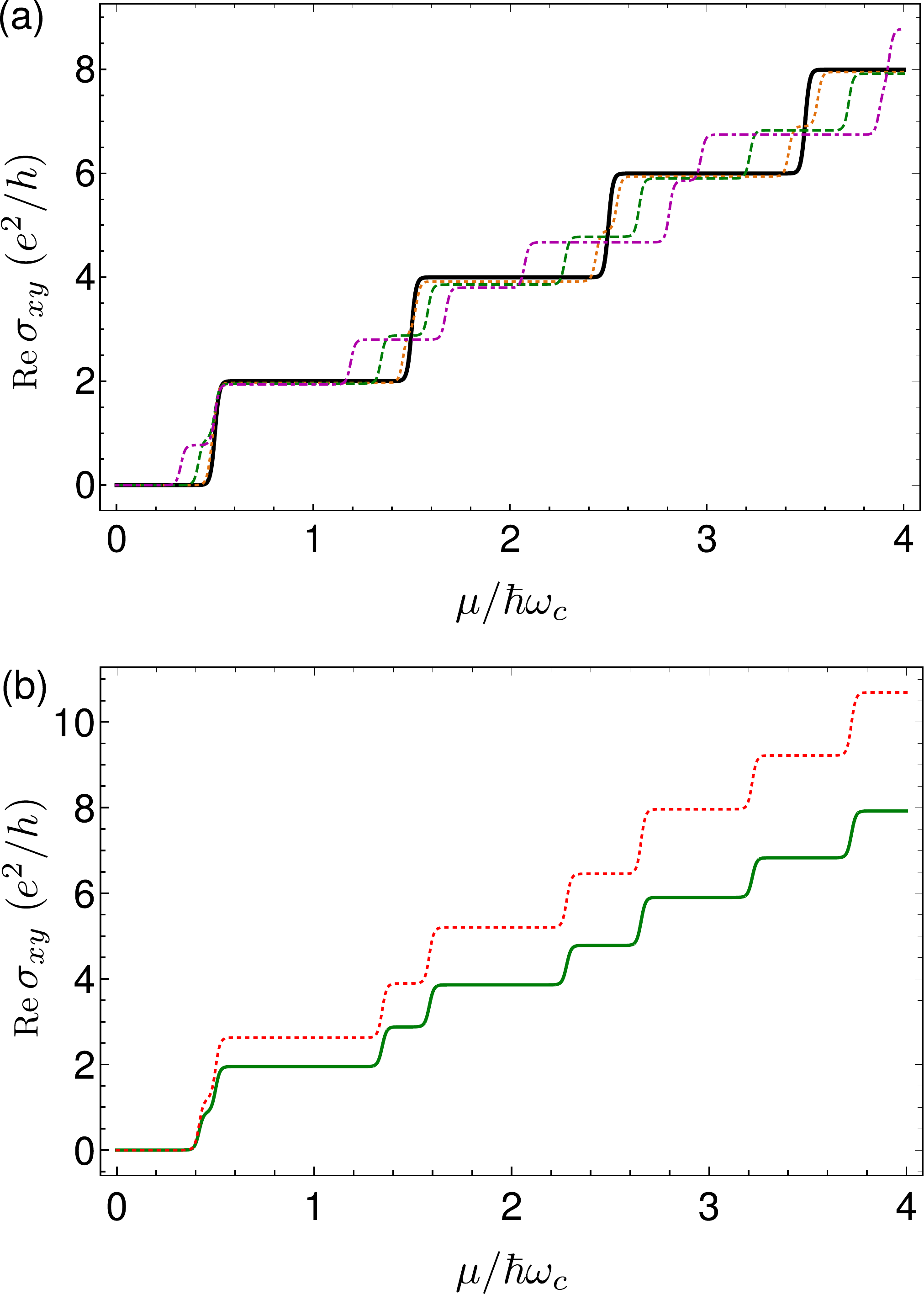} % first figure itself
  	  \caption{(a) Static transverse conductivity, $ \textnormal{Re}\, \sigma_{xy}(0)$,
  	  as a function of the normalized chemical potential $\mu/(\hbar\omega_c)$.
  	  Each curve corresponds to a different value of the parameter $\lambda_B/(\hbar\omega_c)$: 0 (black solid line), 0.1 (orange dotted line), 0.2 (violet dot-dashed  line), 0.3 (green dashed line).
  	  The effective level broadening and temperature are \textcolor{black}{$\Gamma/(\hbar \omega_c)=0.05$} and 
  	  $k_B T/(\hbar \omega_c) =0.01$ respectively.
  	  (b) Transverse photoconductivity, $ \textnormal{Re}\, \sigma_{xy}(\Omega)$, as a function of $\mu/(\hbar \omega_c)$ \textcolor{black}{for $\Omega/\omega_c = 0.5$ (red dotted line)}. The green solid line corresponds to the static transverse conductivity. We have considered $\lambda_B/(\hbar \omega_c) = 0.3$ and employed the same temperature and level broadening as in panel (a).
  	   }\label{fig:f1}
\end{figure}

\subparagraph{Static limit, $\{\xi, \Omega \} \rightarrow 0$.}
We first analyze the static transverse conductivity, $\sigma_{xy}(0)$, using Eq. \eqref{eq:HallC}.%
For $\xi \rightarrow 0$, the Floquet-Landau energies and eigenstates \eqref{quasi} and \eqref{evfull} reduce
to Eqs. \eqref{energy0} and \eqref{ev0}. The only $\Omega$ dependency left is in the denominator
of Kubo formula and vanishes trivially at $\Omega = 0$.
The resulting conductivity is shown in  Fig. \ref{fig:f1} (a)
for several values of $\lambda_B/(\hbar \omega_c)$, parameter measuring the SO coupling strength per magnetic length normalized to the cyclotron energy \footnote{For simplicity, we
assume a vanishing Zeeman field, $\Delta = 0$, taking into consideration a non-vanishing Zeeman field being straightforward.}. 
For $\lambda_B/(\hbar \omega_c) = 0$ (vanishing Rashba SO interaction), the 
 conductivity shows the expected sequence of quantized Hall plateaus appearing at integer units of the conductance quantum ($e^2/h$). 
For $\lambda_B/(\hbar \omega_c) \neq 0$ (non-vanishing Rashba SO coupling), we
observe the appearance of additional plateaus at odd integer units of $e^2/h$. 
These plateaus result from the SO split-Landau levels and 
possess different widths since in the presence of SO interaction, the resulting Landau levels
are no longer equidistant in energy.
When higher Landau levels are populated for larger values of the chemical potential, the conductance shows small
deviations from
the expected quantization. These small deviations occur when
the effect of the SO interaction in the level splitting starts to be relevant compared to the kinetic term
in the Hamiltonian. They have already been
pointed out in previous works when the Hamiltonian combines quadratic and linear terms in the momentum \cite{Carbotte2014} 
and its origin attributed
to the perturbative formulation of transport theory used here\cite{Hernangomez2014}.
\subparagraph{Dynamic limit, $\{\xi, \Omega\} \neq 0$.}
We now study the photoconductivity from Eq. \eqref{eq:HallC} at non-zero radiation frequency.
Fig. \ref{fig:f1} (b) shows the photoconductivity as a function of the normalized chemical potential, $\mu/(\hbar \omega_c)$.
 We choose a value
for the SO interaction of $\lambda_B/(\hbar \omega_c) = 0.3$, which corresponds to typical values for BiSb surface gases, and
select a photon frequency non-resonant with the transition energy between Floquet-Landau levels.
We find that $\textnormal{Re} \,\sigma_{xy}(\Omega)$ preserves a plateau-like structure 
but  no longer quantized in
integer units of $e^2/h$. The reasonable robustness of the step structure is 
reminescent to the behavior of the optical conductivity in the absence of SO interaction\cite{Morimoto2009}. This behavior 
can be expected as the radiation field only renormalizes the magnetically-dressed SO interaction and Zeeman terms
in the photo-induced regime under consideration.

\begin{figure}
  	  \centering
  	  \includegraphics[width=0.45\textwidth]{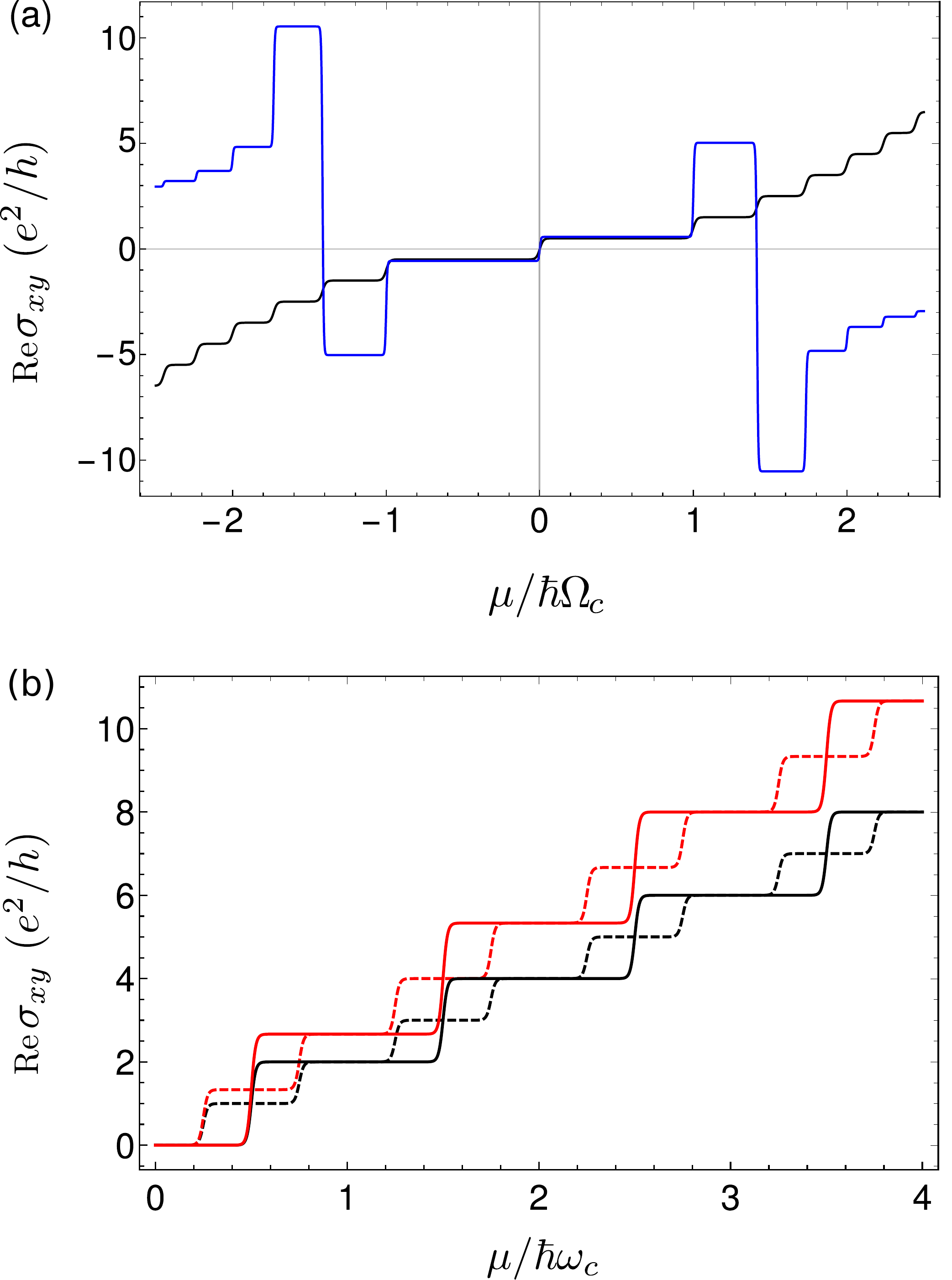} 
  	  \caption{(a) Transverse conductivity, $ \textnormal{Re}\, \sigma_{xy}(\Omega)$, as a function of $\mu/(\hbar \Omega_c)$ obtained in the limit of strong SO interaction, $\lambda_B/(\hbar \omega_c) \gg 1$. Here $\Omega_c = \lambda_B/\hbar$ represents the SO-dependent characteristic frequency. 
  	   Representative results for the static (black line) and
  	    photoinduced cases \textcolor{black}{($ \Omega/\Omega_c = 0.5$, red line)} are shown. 
  	    (b) Transverse conductivity, $\textnormal{Re}\, \sigma_{xy}(\Omega)$, as function of $\mu/\hbar \omega_c$ obtained in the limit of vanishing SO interaction, $\lambda_B/(\hbar \omega_c) \ll 1$. 
  	  %
%   	  % 
   	  Representative results for the static (black curve) and photoinduced cases (\textcolor{black}{$\omega/\omega_c= 0.5$,} red line) are shown. Solid and dashed lines correspond, respectively, to $\delta/(\hbar \omega_c) = 0$ and $\delta/(\hbar \omega_c) \simeq 1.2$.
   	    The effective level broadening and the temperature are chosen as in Fig. \ref{fig:f1} both for panels (a) and (b). 
  	  }\label{fig:f2}
\end{figure}

\subparagraph{Quasi-relativistic limit, $\lambda_B/\hbar\omega_c \gg 1$.}
We now consider Eq. \eqref{eq:HallC} when the energy scale 
associated to the SO interaction dominates over the cyclotron energy (formally, this is equivalent to 
consider the formal limits $\Delta \rightarrow 0$ and $m^\ast \rightarrow +\infty$ simultaneously).
 In this limit, 
 %the energy scale related to the Rashba SO interaction dominates and 
 the energy spectrum  is gapless in the absence of external magnetic and electric fields. 
 In the presence of magnetic field but no coupling to the radiation field, the 
 Landau levels reduce to the well-known expression $\varepsilon_{sm} = s \hbar \Omega_c \sqrt{m}$.
 Here, $\Omega_c = \lambda_B / \hbar = \lambda \sqrt{2} /l_B$ is a SO-dependent
 characteristic frequency analogous to the graphene cyclotron frequency\cite{Gusynin2005, Lopez2015} with $\lambda$ playing 
 a role analogous to the Fermi velocity.

 Fig. \ref{fig:f2} (a) displays the transverse conductivity in this limit (see Appendix \ref{app:details_conductivity} for analytical details)
 for the static (blue curve) and dynamic (red curve) cases
 as a function of the chemical potential. % $\mu/(\hbar \Omega_c)$.
 In the absence of coupling to the radiation field, 
 we recover the well-known half-integer quantization of
 the transverse conductivity that occurs for single Dirac
 cones at the surface of topological insulators\cite{Burkov2011, Hernangomez2014, Konig2014} (note that in graphene, due to the combined effect
 of spin and pseudo-spin degeneracies the conductivity is four times bigger\cite{Gusynin2005, Morimoto2009}).
 The radiation field strongly modifies the form of the transverse conductivity while preserving
 particle-hole symmetry [in other words, the conductivity is still an odd function of $\mu/(\hbar \Omega_c)$]. Similar to the static case
 a step-like structure
 is still preserved for the first steps, however for larger values of the chemical potential the conductivity is no longer a monotonic function of $\mu$.
 Compared to Fig. \ref{fig:f1} (b), $\sigma_{xy}(\Omega)$
has a more complex structure.
The low energy resonances are
associated to allowed transitions between quasi-relativistic Landau levels with $s = s'$ and $|n - n'| = 1$.
They involve only the electron energy sector. At higher energy, the resonances result from transitions that
involve both electron and hole energy sectors, characterized by $s \neq s'$ and $|n - n'| = 1$.

\subparagraph{Non-relativistic limit, $\lambda_B/\hbar\omega_c \ll 1$.}
We finally consider the opposite limit, which corresponds to neglect the leading
SO interaction contributions
to the static Hamiltonian.
It is easy to verify (see details in Appendix \ref{app:details_conductivity})
that the transverse conductivity reduces to
\begin{equation}
    \sigma_{xy} (\Omega) \simeq \frac{e^2}{h} \sum_{m = 0}^{+\infty}\sum_{\sigma} \frac{\omega_c^2}{\omega_c^2-(\Omega+i\Gamma/\hbar)^2}n_\textnormal{F}(\varepsilon_{\sigma m}),
\end{equation}
where $\varepsilon_{\sigma m} = \hbar \omega_c\left[ m + {1}/{2} + {\sigma} {\Delta}/{(2\hbar\omega_c)} \right]$.
We display in Fig. \ref{fig:f2} (b) traces
for the static (black trace) and dynamic (red trace) conductivity as a function of the normalized chemical potential, $\mu/(\hbar \omega_c)$.
The solid / dashed lines
correspond, respectively, to absence [$\delta/(\hbar \omega_c) = 0$, continuous] or presence [$\delta/(\hbar \omega_c) \simeq 1.2$, dashed] of 
Zeeman coupling to the static magnetic field.
As expected, including Zeeman interaction results in the appearance
of additional plateaus,  all of them having the same
width contrary to Fig. \ref{fig:f1} (a).
As reported previously\cite{Morimoto2009}, we obtain a robust step-like structure
under illumination departing from the exact quantization
in units of $e^2/h$. This is similar to the case with SO interaction shown in Fig. \ref{fig:f1} (b) suggesting
that the renormalization of the eigenstates due to the radiation has a subdominant role in the photoconductivity properties in this regime and that the impact
of the SO interaction in the photo-induced conductivity is, at most, of quantitative nature.

\section{Conclusion and outlook}\label{sec:conclusion}
In conclusion, we have studied photoinduced phenomena in 2DEG under perpendicular magnetic fields with Rashba SO interaction.
Our work provides perturbative analytical expressions for physical observables valid
in the THz / low infrared regime.
We have first presented the photoinduced modulation to the Landau
energy levels in the presence of a continuous and periodic radiation field.
Using the Floquet-Landau states, we have considered the dynamical features
on the spin polarization and the autocorrelation function.
Assuming that the initially prepared state is a coherent state, which
possesses a static finite spin polarization, we have shown that the exchange of angular momentum between the charge
carriers and the radiation field mediated by SO interaction (which
couples spin and orbital degrees of freedom) \textcolor{black}{produces oscillations 
of the average spin polarization as a function of the magnetic field strength.}
For large enough values of the coherent state parameter $\alpha$ we have
demonstrated that the time evolution of the spin polarization
has periodic beating patterns due to dynamical localization and
interference of Floquet-Landau levels.
\textcolor{black}{These} effects can be correlated \textcolor{black}{with} fractional
revivals in the autocorrelation function.
Using the Kubo formalism, we have computed
the transverse photoconductivity of the system. We have found  that the SO interaction does not drastically
change the resonant structure of the transverse photoconductivity (compared to the 2DEG under illumination). 
When considered as a function of the chemical potential, the photoconductivity shows non-quantized plateaus of different width due to the combined effect of the radiation field and the SO interaction.
This is different from the photoconductivity of graphene or topological
insulator surfaces characterized by (static) linear dispersion relation,
as the latter shows resonances associated to additional allowed transitions between Landau levels.
We have verified that our analytical expression yields well-known results in the special limits of vanishing SO interaction [$\lambda_B/(\hbar \omega_c) \ll 1$] and that we recover results for systems with a ``quasi-relativistic'' dispersion relation whenever
the SO interaction strongly dominates [$\lambda_B/(\hbar \omega_c) \gg 1$].
Finally, one key point to be addressed is the experimental feasibility of our proposal. 
As shown in Secs. \ref{sec3} and \ref{sec4}, we have considered parameter estimates for static magnetic field, Rashba SO interaction as well as the amplitude and frequency of the radiation field compatible with achievable experimental systems (BiSb or InSb surface gases irradiated by THz / infrared radiation). 
Now, the effects of electromagnetic dressing of the energy bands can be experimentally explored by studying the optical response of the system in a pump-probe geometry made of these materials. 
In this case, a sample is excited by a continuous-wave highly intense laser
(pump) while the second pulse (probe) is used for characterization of the excited states of the light-matter coupled system\cite{Yudin2016}. Thus, within the possible experimental techniques to be considered we can suggest the time-resolved angle resolved photoemission spectroscopy (ARPES), which has been considered previously for topological insulators in Refs. \onlinecite{Wang2013,Mahmood2016} and extended to deal with the spin-resolved polarization (the so-called SARPES technique) in Ref. \onlinecite{doi:10.1021/nl504693u}.
In addition, a modification of the scheme presented in Ref. \onlinecite{Bayer_2019} where they combine bichromatic polarization pulse shaping with photoelectron imaging tomography for time-resolved spatial imaging of an ultrafast SO-split wave packet could afford a suitable experimental tool for testing our results on the time-dependent modulation of the coherent state wave packet spin polarization. We note
that if one would be interested in experimentally addressing some ``non-universal'' features of the photoconductivity in graphene-like systems effects of doping and finite temperature would need to be included\cite{Mak2008}.

\begin{acknowledgments}
A. L. acknowledges fruitful discussions with Benjamin Santos.
Funding by CEDIA via Project CEPRA-XII-2018-06 "Espectroscop\'ia Mec\'anica: transporte e interacci\'on materia-radiaci\'on" and the German Research Foundation (DFG) through the Collaborative Research Center, Project ID 314695032 SFB 1277 (projects A03, B01) is gratefully acknowledged.

\end{acknowledgments}

%%%%%%%%%%%%%%%%%%%%%%%%%%%%%%%%%%%%%%%%%%%%%%%%%%%%%%%%%
%%%%%%%%%%%%%%%%%%%%%%%%%%%%%%%%%%%%%%%%%%%%%%%%%%%%%%%%%

\appendix

%%%%%%%%%%%%%%%%%%%%%%%%%%%%%%%%%%%%%%%%%%%%%%%%%%%%%%%%%
\section{Perturbative calculation of the effective Floquet  Hamiltonian}\label{app:perturbative_terms}
%%%%%%%%%%%%%%%%%%%%%%%%%%%%%%%%%%%%%%%%%%%%%
In order to obtain the effective Floquet Hamiltonian given in Eq. (\ref{hfin}), we evaluate the following expression
\begin{equation}\label{eq:floquet_full}
H=e^{-i\kappa/2 I_+}H_Fe^{i\kappa/2 I_+},
\end{equation}
with the operators $I_{\pm}$ (resp. Hermitian and anti-Hermitian) defined as
\begin{equation}
I_{\pm}={a}^\dagger\sigma_- \pm {a}\sigma_+. 
\end{equation}
Using the Baker-Campbell-Hausdorff formula we have
\begin{equation}\label{eq:floquet_BCH_expansion}
H = H_F-\dfrac{i\kappa}{2}[I_+,H_F]+\dfrac{1}{2!}\left(\dfrac{i\kappa}{2}\right)^2[I_+,[I_+,H_F]]+\dots
\end{equation}
To leading order in $\kappa$, only the first commutator needs to be evaluated. We find
\begin{equation}\label{eq:con1}
[I_+,H_F]=-\frac{\delta}{2}[I_+,\sigma_z]+i\lambda_B[I_+,I_-]-\xi[I_+,\sigma_y],
\end{equation} 
where we have used that $[I_+,N_a]=0$. It is a straightforward task to evaluate the commutators in Eq. \eqref{eq:con1}, we obtain $[I_+,\sigma_z]=I_-$, $[I_+,I_-]=2N_a\sigma_z$ and $[I_+,\sigma_y]=i(a+a^\dagger)\sigma_z$.

Upon substitution of these results in Eq. \eqref{eq:floquet_BCH_expansion}, we get
\begin{align}
H_F^\textnormal{eff}  &=\hbar\omega_- N_a-\Big( \frac{\delta-2\kappa\lambda_B N_a}{2}\Big)\sigma_z \nonumber \\ 
&+ i\Big(\lambda_B+\frac{\alpha\delta}{2}\Big) I_- - \frac{\kappa\xi}{2}(a^\dagger+a)\sigma_z-\xi\sigma_y.
\end{align}
We now define the shifted ladder operators $c=a-\gamma$, with $\gamma=2\xi/(2\lambda_B+\kappa\delta)$. With this definition, the following relations are satisfied 
\begin{subequations}
\begin{align}
N_a& = N_c+\gamma(c^\dagger+c)+\gamma^2,\\
a^\dagger+a &= c^\dagger+c+2\gamma,
\end{align}
\end{subequations}
and allow us in turn to write the perturbative Hamiltonian as
 \begin{align}
H_F^\textnormal{eff}  &=\hbar\omega_- N_c+\gamma\hbar\omega_-(c^\dagger+c)+\gamma^2\hbar\omega_- \notag \\
 &-\Big( \frac{\delta-2\kappa\lambda_B N_c}{2}\Big)\sigma_z+\kappa\lambda_B\gamma(c^\dagger+c)+\kappa\lambda_B\gamma^2 \notag \\ 
 &+i\Big(\lambda_B+\frac{\kappa\delta}{2}\Big) (c^\dagger\sigma_--c\sigma_+) \nonumber\\
&-\frac{\kappa\xi}{2}(c+c^\dagger+2\gamma)\sigma_z.
\end{align}
Our effective Floquet Hamiltonian
\begin{align}
H_F^\textnormal{eff}&= \hbar\omega_- N_c-\Big( \frac{\delta-2\kappa\lambda_B N_c}{2}\Big)\sigma_z  \notag \\
& +i\Big(\lambda_B+\frac{\kappa\delta}{2}\Big) (c^\dagger\sigma_--c\sigma_+),
\end{align}
 where we have neglected higher order terms in $\gamma$ and $\xi$
  \begin{align}
\Delta H&=\frac{2\gamma(\kappa\lambda_B+\hbar\omega_-)-\kappa\xi\sigma_z}{2}(c^\dagger+c) \notag \\ &+\gamma^2(\kappa\lambda_B+\hbar\omega_-)-\kappa\gamma\xi\sigma_z.
\end{align}

%%%%%%%%%%%%%%%%%%%%%%%%%%%%%%%%%%%%%%%%%%%%%%%%%%%%%%%%%%%
\section{Calculation details of the spin polarization}\label{app:details_spinpolarization}

We want to evaluate the mean spin polarization
\begin{equation}\label{sigmaz2}
\langle\sigma_z\rangle= \dfrac{1}{T} \int_0^{T} \textnormal{d}t \,\sigma_z(t), 
\end{equation}
where $\sigma_z(t) = \bra{\Psi(0)}U^\dagger(t)\sigma_zU(t)\ket{\Psi(0)}$, for the initial state of the system given by the coherent state $\ket{\alpha}=\ket{\Psi(0)}$,
\begin{equation}\label{eq:coherent2}
\ket{\alpha}=e^{-\frac{\lvert\alpha\rvert^2}{2}}\Big(\ket{\varphi_0}+\dfrac{1}{\sqrt{2}} \sum_{n=1}^{+\infty} \sum_{s} \dfrac{\alpha^n}{\sqrt{n!}}\ket{\varphi_{sn}}\Big).
\end{equation}
We begin by transforming the coherent state to the eigenbasis of the effective Floquet Hamiltonian
\begin{equation}\label{coherent3}
\ket{\alpha}=\ket{\psi_0}\bra{\psi_0}\alpha\rangle+\sum_{n=1}^{+\infty} \sum_{s}\ket{\psi_{sn}}\bra{\psi_{sn}}\alpha\rangle,
\end{equation}
with expansion coefficients
\begin{subequations}
\begin{align}
\bra{\psi_0}\alpha\rangle &= e^{-\frac{\lvert\alpha\rvert^2}{2}},\\
\bra{\psi_{sn}}\alpha\rangle &= e^{-\frac{\lvert\alpha\rvert^2}{2}}\frac{\alpha^n}{\sqrt{n!}}b_{-s,n}.
\end{align}
\end{subequations}
Using the approximation $H_F\simeq H$, we get
\begin{align}
e^{-iH t/\hbar}&\ket{\alpha}=e^{-\frac{\lvert\alpha\rvert^2}{2}}\Big[e^{-i\delta t/2\hbar}\ket{\psi_0} \notag \\ &+ \sum_{m=1}^{+\infty} \sum_{s}\frac{\alpha^m}{\sqrt{m!}}b_{-s,m}e^{-i(s\varepsilon_m+m\hbar\omega_-)t/\hbar}\ket{\psi_{sn}}\Big]
\end{align}
and the results
\begin{subequations}
\begin{eqnarray}
\bra{\psi_0}\sigma_z\ket{\psi_0}&=&-1, \notag \\
\bra{\psi_{s'm'}}\sigma_z\ket{\psi_{sm}}&=&(ss'b_{-s'm'}b_{-sm}-b_{s'm'}b_{sm})\delta_{mm'} \notag,
\end{eqnarray}
\end{subequations}
we obtain
\begin{align}
\sigma_z(t)&=-e^{-\lvert\alpha\rvert^2} 
 \Big[1+\sum_{m=1}^{+\infty}\sum_{ss'}\dfrac{|\alpha|^{2m}}{m!}b_{-s',m}b_{-s,m} \notag \\
  & \times e^{i(s'-s)\varepsilon_m} (b_{s'm}b_{sm}-ss'b_{-s'm}b_{-sm})\Big].   
\end{align}
Performing the double sum in $s$ and $s'$ we get
\begin{eqnarray}
\sigma_z(t)&&=-e^{-\lvert\alpha\rvert^2} 
 \Bigg\{1+\sum_{m=1}^{+\infty}\dfrac{|\alpha|^{2m}}{m!}\Big[\Big(\frac{\tilde{\Delta}_m}{\varepsilon_m}\Big)^2 \notag \\ &&+\Big(1-\frac{\tilde{\Delta}_m^2}{\varepsilon_m^2}\Big)\cos \left(\dfrac{2\varepsilon_mt }{\hbar} \right)\Big]\Bigg\}.\nonumber 
\end{eqnarray}
It can be easily checked that this expression reduces to the result obtained for the pseudospin polarization in irradiated graphene under perpendicular magnetic fields from Ref. \onlinecite{Lopez2015} in the limit $\lambda_B / (\hbar \omega_c) \gg 1$. Finally, upon of integration of $\sigma_z(t)$ over one period of oscillation of the radiation field, $T = 2\pi/\Omega$, we obtain the result quoted in Eq. \eqref{eq:spin_polarization}.

\section{Calculation details of the autocorrelation function}\label{app:detains_autocorrelation}
We want to evaluate the autocorrelation function $C(t)= \braket{\Psi(0)|\Psi(t)}$ for the 
initial state of the system given by the coherent state \eqref{coherent}.
We begin by writing the states in Eq. \eqref{ev0} in terms of the Floquet basis \eqref{evfull} for which the time evolution is trivial. We find $|\alpha(t)\rangle$ to be
\begin{align}
|\alpha(&t) \rangle = {\rm e}^{-\frac{|\alpha|^2}{2}} \Big[e^{-i\delta t/2\hbar} |\psi_0 \rangle + \dfrac{1}{\sqrt{2}} \sum_{m=1}^{+\infty} \sum_{ss'} \dfrac{\alpha^m}{m!} \notag  \\ &\times  (c_{-sm} b_{-s'm} + s s' c_{sm} b_{s'm}) {\rm e}^{-i (s' \varepsilon_m  + m\hbar\omega_-)t/\hbar}|\psi_{sm}\rangle \Big].
\end{align}
Using Eq. \eqref{coherent} and taking into account the orthogonality properties of the Floquet eigenstates we obtain for $C(t)$ 
\begin{align}
&C_\alpha(t) = {\rm e}^{-\frac{|\alpha|^2}{2}} \Big\{ 1 + \dfrac{1}{2} \sum_{m=1}^{+\infty} \dfrac{|\alpha|^{2m}}{m!} \sum_{s}  {\rm e}^{-i s \varepsilon_m t/\hbar}\Big[ (c_{sm} b_{sm})^2  \notag \\ & +  (c_{-sm} b_{sm})^2 + (c_{sm} b_{-sm})^2 +(c_{-sm} b_{-sm})^2 \notag \\ & + 2 (c_{sm} b_{sm} + c_{-sm} b_{-sm})(c_{sm} b_{-sm} - c_{-sm} b_{sm})
\Big]
\Big\}.
\end{align}
Performing the summation over $s$ and using Eqs. \eqref{cs} and \eqref{bs} we get the expression quoted in Eq. \eqref{eq:autocorrelation}.

%%%%%%%%%%%%%%%%%%%%%%%%%%%%%%%%%%%%%%%%%%%%%%%%%%%%%%%%%%%%%
\section{Calculation details of the photoconductivity}\label{app:details_conductivity}

\subparagraph{Generalities.}
We compute the photoconductivity for the Floquet system using the Kubo formula
\begin{align}
&\sigma_{\mu\nu}(\Omega)=\dfrac{i e^2 \hbar}{L^2}\sum_{\varepsilon_a, \varepsilon_b} \frac{n_\textnormal{F}(\varepsilon_b) - n_\textnormal{F}(\varepsilon_a)}{\varepsilon_b-\varepsilon_a} \notag  \\ &\times\Bigg[
\frac{v_\mu^{ab}v_\nu^{ba}}{\varepsilon_b-\varepsilon_a-(\hbar\Omega+i\Gamma)}  -\frac{v_\nu^{ab}v_\mu^{ba}}{\varepsilon_b-\varepsilon_a+\hbar\Omega+i\Gamma}\Bigg],\label{eq:kubo}
\end{align} 
where the Fermi-Dirac distribution function, $n_\textnormal{F}(E)$, is given by Eq. \eqref{eq:FD}. In Eq. \eqref{eq:kubo}, $v^{ab}_\mu$ represent the matrix elements of the velocity operator, $v^{ab}_\mu := \langle \psi_a|  v_\mu| \psi_b \rangle$, where the states $| \psi_a \rangle$ are given \textcolor{black}{by Eq. \eqref{evfull}}.
We also consider an effective energy level broadening due to scattering with impurities with the phenomenological parameter $\Gamma$. This parameter is considered to be 
substantially smaller than the photon energy $\Gamma\ll\hbar\Omega$, which is valid for THz radiation and low impurity concentration.
Note that in the standard Kubo formula \eqref{eq:kubo}, there are no contributions
 from the Floquet replicas, and therefore, the static limit of the conductivity
 can be obtained from the limit $\Omega \rightarrow 0$.

The components of the velocity operator
are easily obtained from the equation of motion
$v_\mu=[r_\mu, \textcolor{black}{H(t)}]/(i\hbar)$, with $r_\mu$ ($\mu=x,y)$ being the components of the position operator and $H(t)$ given by the full
Hamiltonian, $H(t) = H_0 + V(t) $.
The commutator can be calculated trivially and the velocity thus reads
\begin{equation}
	v_\mu = \frac{1}{m^\ast} \pi_\mu + \lambda \epsilon_{\mu \nu z}  \sigma_\nu,
\end{equation}
where $\epsilon_{\mu\nu\lambda}$ is the Levi-Civita symbol.
We transform to the Floquet basis $\{ c^\dagger,c\}$
to find 
\begin{subequations}
\begin{align}
	v_x &= \dfrac{\hbar}{\sqrt{2}l_B m^*} (c^\dagger +c) + \lambda \sigma_y, \\
	v_y &= \dfrac{\hbar}{\sqrt{2} i l_B m^*} (c^\dagger - c) - \lambda \sigma_x,
\end{align}
\end{subequations}
where we neglected additive terms proportional to $\gamma$ [as they correspond to higher order perturbation terms for $\sigma_{xy}(\Omega)$].
We compute the matrix elements $v_\mu^{ab}$ using the eigenstates in Eq. \eqref{evfull}. As expected, the conductivity comes from the off-diagonal terms
of the velocity operator, \textit{i.e.} $v_\mu^{m,m'} = v_\mu^m \delta_{m,m' \pm 1}$, that couple Floquet-Landau levels with quantum number $m$ differing by one.
The explicit calculation of the velocity matrix elements shows this feature, \textit{e.g.} the matrix element for the $x$ component is given by
\begin{align}
     v_x^{m,m'}&=\Big[\sqrt{m'} B^{ss'}_{mm'} +ss'\sqrt{m'+1} B^{-s-s'}_{mm'}\Big]\delta_{m,m'+1}\\
     &+\Big[\sqrt{m'-1} B^{ss'}_{mm'} +ss'\sqrt{m'} B^{-s-s'}_{mm'}\Big]\delta_{m,m'-1}\notag\\
     &+\frac{\lambda_B}{\hbar\omega_c} \Big[s' B^{s-s'}_{mm'} \delta_{m,m'-1} + s B^{-ss'}_{mm'} \delta_{m,m'+1}\Big]. \notag
\end{align}
Here, we defined the combination of wavefunction weights $B^{ss'}_{mm'} = b_{sm}b_{s'm'}$. Using that $v_x^{ab}v_y^{ba}=-v_x^{ba}v_y^{ab}$, and taking into account the cancellation of terms due to products of delta functions, we obtain Eq. \eqref{eq:HallC} of the main text.

%%%%%%%%%%%%%%%%%%%%%%%%%%%%%%%%%%%%%%%%%%%%%%%%%%%%%%%%%%%
\subparagraph*{Calculation details for the $\lambda_B / (\hbar \omega_c) \gg 1$ limit.}
When the energy scale related to the Rashba SO interaction dominates over the cyclotron energy,  $\lambda_B / \hbar \omega_c \gg 1$, the Floquet-Landau spectrum in Eq. \eqref{quasi} reduces to
\begin{equation}\label{en_graph}
    \varepsilon_{sm} = s\lambda_B \sqrt{m + m^2 \kappa^2}.
\end{equation}
We recognize that $\lambda_B$ plays the role of a ``graphene-like'' cyclotron energy with a SO defined cyclotron frequency $\Omega_c := \lambda_B/\hbar = \lambda \sqrt{2}/l_B$.
We consider for simplicity the limit of weak-coupling to the radiation field, $\kappa \ll 1$ and $\varepsilon_{sm} \simeq  s\hbar \Omega_c \sqrt{m}$. Using that $b_{sm}\simeq 1/\sqrt{2}$ and considering
only the dominating term proportional to $\lambda_B$ in Eq. \eqref{eq:kubo} we find
\begin{equation}
\sigma_{xy}(\Omega) \simeq \dfrac{e^2}{4 h} \sum_{m=0}^{+\infty} \sum_{ss'} \dfrac{n_\textnormal{{F}} (\varepsilon_{s'm+1}) - n_\textnormal{{F}} (\varepsilon_{sm})}{(s'\sqrt{m+1} - s\sqrt{m})^2 - \left(\frac{\Omega}{\Omega_c} + i\frac{\Gamma}{\hbar \Omega_c}\right)^2}.
\end{equation}
In the static limit ($\Omega \rightarrow 0$) with $\Gamma = 0$ this expression reduces to
the result presented in, \textit{i.e.}, Refs. [\onlinecite{Carbotte2014, Hernangomez2014, Konig2014}] for single surfaces of topological
insulators under perpendicular magnetic field.

This result is not surprising, as we observe that the structure of the Floquet-Landau eigenstates greatly simplifies when SO coupling becomes the dominant energy scale compared to the cyclotron energy. 
This results in a loss of ``memory'' of the interaction with the radiation field in the Floquet-dressed velocity matrix elements. Consequently, the radiation only appears in the denominator of Kubo formula, as in the standard scenario of linear response optical conductivity \onlinecite{Carbotte2014}.

\subparagraph*{Calculation details of the  $\lambda_B /(\hbar \omega_c) \ll 1$  limit.}

In the limit of vanishing SO interaction, $\lambda_B/\hbar\omega_c \ll 1$, 
the spin, $\sigma = \pm$ becomes a good quantum number (instead of the SO quantum number, $s$).
It is easy to check that we can reintroduce the picture
of Zeeman-split spin polarized Landau levels provided that
we map the energy levels as $(m,s) \rightarrow (m_\sigma, \sigma)$
where $m_\sigma = m - (1 + \sigma)/2$. 
The eigenenergies \eqref{quasi} then reduce to
\begin{equation}\label{en_2deg}
    \varepsilon_{\sigma m} = \hbar \omega_c\left( m + \frac{1}{2} + \frac{\sigma}{2} \frac{\Delta}{\hbar\omega_c} \right),
\end{equation}
where we also assume weak-coupling to the radiation field.
Using  that the coefficients $b_{sm}\simeq \sqrt{(1-s)/2}$,
the transverse conductivity \eqref{eq:HallC} reduces to
\begin{equation}
    \sigma_{xy} (\Omega) \simeq \frac{e^2}{h} \sum_{m = 0}^{+\infty}\sum_{\sigma} \frac{\omega_c^2}{\omega_c^2-(\Omega+i\Gamma/\hbar)^2}n_\textnormal{F}(\varepsilon_{\sigma m}),
\end{equation}
a result given in Ref. [\onlinecite{Morimoto2009}] and also reproduced in the main text.
Similar to the case $\lambda_B / \hbar \omega_c \gg 1$, we observe that if the SO coupling vanishes the structure of the Floquet-Landau states becomes trivial. In addition, each state has only $+$ or $-$ component). Consequently, 
the coupling to the radiation field disappears  from the velocity matrix elements (as it also disappears from the Hamiltonian) and the photon energy only enters in the denominator of 
the Kubo formula.
%

% \bibliography{biblio}

\begin{thebibliography}{57}%
\makeatletter
\providecommand \@ifxundefined [1]{%
 \@ifx{#1\undefined}
}%
\providecommand \@ifnum [1]{%
 \ifnum #1\expandafter \@firstoftwo
 \else \expandafter \@secondoftwo
 \fi
}%
\providecommand \@ifx [1]{%
 \ifx #1\expandafter \@firstoftwo
 \else \expandafter \@secondoftwo
 \fi
}%
\providecommand \natexlab [1]{#1}%
\providecommand \enquote  [1]{``#1''}%
\providecommand \bibnamefont  [1]{#1}%
\providecommand \bibfnamefont [1]{#1}%
\providecommand \citenamefont [1]{#1}%
\providecommand \href@noop [0]{\@secondoftwo}%
\providecommand \href [0]{\begingroup \@sanitize@url \@href}%
\providecommand \@href[1]{\@@startlink{#1}\@@href}%
\providecommand \@@href[1]{\endgroup#1\@@endlink}%
\providecommand \@sanitize@url [0]{\catcode `\\12\catcode `\$12\catcode
  `\&12\catcode `\#12\catcode `\^12\catcode `\_12\catcode `\%12\relax}%
\providecommand \@@startlink[1]{}%
\providecommand \@@endlink[0]{}%
\providecommand \url  [0]{\begingroup\@sanitize@url \@url }%
\providecommand \@url [1]{\endgroup\@href {#1}{\urlprefix }}%
\providecommand \urlprefix  [0]{URL }%
\providecommand \Eprint [0]{\href }%
\providecommand \doibase [0]{http://dx.doi.org/}%
\providecommand \selectlanguage [0]{\@gobble}%
\providecommand \bibinfo  [0]{\@secondoftwo}%
\providecommand \bibfield  [0]{\@secondoftwo}%
\providecommand \translation [1]{[#1]}%
\providecommand \BibitemOpen [0]{}%
\providecommand \bibitemStop [0]{}%
\providecommand \bibitemNoStop [0]{.\EOS\space}%
\providecommand \EOS [0]{\spacefactor3000\relax}%
\providecommand \BibitemShut  [1]{\csname bibitem#1\endcsname}%
\let\auto@bib@innerbib\@empty
%</preamble>
\bibitem [{\citenamefont {Lindner}\ \emph {et~al.}(2011)\citenamefont
  {Lindner}, \citenamefont {Refael},\ and\ \citenamefont
  {Galitski}}]{NP2011Lindner}%
  \BibitemOpen
  \bibfield  {author} {\bibinfo {author} {\bibfnamefont {N.~H.}\ \bibnamefont
  {Lindner}}, \bibinfo {author} {\bibfnamefont {G.}~\bibnamefont {Refael}}, \
  and\ \bibinfo {author} {\bibfnamefont {V.}~\bibnamefont {Galitski}},\ }\href
  {\doibase 10.1038/nphys1926} {\bibfield  {journal} {\bibinfo  {journal}
  {Nature Physics}\ }\textbf {\bibinfo {volume} {7}},\ \bibinfo {pages} {1745}
  (\bibinfo {year} {2011})}\BibitemShut {NoStop}%
\bibitem [{\citenamefont {Peng}\ and\ \citenamefont
  {Refael}(2019)}]{PRL2018_Refael}%
  \BibitemOpen
  \bibfield  {author} {\bibinfo {author} {\bibfnamefont {Y.}~\bibnamefont
  {Peng}}\ and\ \bibinfo {author} {\bibfnamefont {G.}~\bibnamefont {Refael}},\
  }\href {\doibase 10.1103/PhysRevLett.123.016806} {\bibfield  {journal}
  {\bibinfo  {journal} {Phys. Rev. Lett.}\ }\textbf {\bibinfo {volume} {123}},\
  \bibinfo {pages} {016806} (\bibinfo {year} {2019})}\BibitemShut {NoStop}%
\bibitem [{\citenamefont {Seetharam}\ \emph {et~al.}(2019)\citenamefont
  {Seetharam}, \citenamefont {Bardyn}, \citenamefont {Lindner}, \citenamefont
  {Rudner},\ and\ \citenamefont {Refael}}]{PRB2018_Lindner}%
  \BibitemOpen
  \bibfield  {author} {\bibinfo {author} {\bibfnamefont {K.~I.}\ \bibnamefont
  {Seetharam}}, \bibinfo {author} {\bibfnamefont {C.-E.}\ \bibnamefont
  {Bardyn}}, \bibinfo {author} {\bibfnamefont {N.~H.}\ \bibnamefont {Lindner}},
  \bibinfo {author} {\bibfnamefont {M.~S.}\ \bibnamefont {Rudner}}, \ and\
  \bibinfo {author} {\bibfnamefont {G.}~\bibnamefont {Refael}},\ }\href
  {\doibase 10.1103/PhysRevB.99.014307} {\bibfield  {journal} {\bibinfo
  {journal} {Phys. Rev. B}\ }\textbf {\bibinfo {volume} {99}},\ \bibinfo
  {pages} {014307} (\bibinfo {year} {2019})}\BibitemShut {NoStop}%
\bibitem [{\citenamefont {Titum}\ \emph {et~al.}(2017)\citenamefont {Titum},
  \citenamefont {Lindner},\ and\ \citenamefont {Refael}}]{PRB2017_Refael}%
  \BibitemOpen
  \bibfield  {author} {\bibinfo {author} {\bibfnamefont {P.}~\bibnamefont
  {Titum}}, \bibinfo {author} {\bibfnamefont {N.~H.}\ \bibnamefont {Lindner}},
  \ and\ \bibinfo {author} {\bibfnamefont {G.}~\bibnamefont {Refael}},\ }\href
  {\doibase 10.1103/PhysRevB.96.054207} {\bibfield  {journal} {\bibinfo
  {journal} {Phys. Rev. B}\ }\textbf {\bibinfo {volume} {96}},\ \bibinfo
  {pages} {054207} (\bibinfo {year} {2017})}\BibitemShut {NoStop}%
\bibitem [{\citenamefont {Esin}\ \emph {et~al.}(2018)\citenamefont {Esin},
  \citenamefont {Rudner}, \citenamefont {Refael},\ and\ \citenamefont
  {Lindner}}]{PRB2018_Lindner_2}%
  \BibitemOpen
  \bibfield  {author} {\bibinfo {author} {\bibfnamefont {I.}~\bibnamefont
  {Esin}}, \bibinfo {author} {\bibfnamefont {M.~S.}\ \bibnamefont {Rudner}},
  \bibinfo {author} {\bibfnamefont {G.}~\bibnamefont {Refael}}, \ and\ \bibinfo
  {author} {\bibfnamefont {N.~H.}\ \bibnamefont {Lindner}},\ }\href {\doibase
  10.1103/PhysRevB.97.245401} {\bibfield  {journal} {\bibinfo  {journal} {Phys.
  Rev. B}\ }\textbf {\bibinfo {volume} {97}},\ \bibinfo {pages} {245401}
  (\bibinfo {year} {2018})}\BibitemShut {NoStop}%
\bibitem [{\citenamefont {Usaj}\ \emph {et~al.}(2014)\citenamefont {Usaj},
  \citenamefont {Perez-Piskunow}, \citenamefont {Foa~Torres},\ and\
  \citenamefont {Balseiro}}]{PRB2014_PerezPiskunow_2}%
  \BibitemOpen
  \bibfield  {author} {\bibinfo {author} {\bibfnamefont {G.}~\bibnamefont
  {Usaj}}, \bibinfo {author} {\bibfnamefont {P.~M.}\ \bibnamefont
  {Perez-Piskunow}}, \bibinfo {author} {\bibfnamefont {L.~E.~F.}\ \bibnamefont
  {Foa~Torres}}, \ and\ \bibinfo {author} {\bibfnamefont {C.~A.}\ \bibnamefont
  {Balseiro}},\ }\href {\doibase 10.1103/PhysRevB.90.115423} {\bibfield
  {journal} {\bibinfo  {journal} {Phys. Rev. B}\ }\textbf {\bibinfo {volume}
  {90}},\ \bibinfo {pages} {115423} (\bibinfo {year} {2014})}\BibitemShut
  {NoStop}%
\bibitem [{\citenamefont {Rudner}\ and\ \citenamefont
  {Lindner}(2020)}]{1909.02008_Lindner}%
  \BibitemOpen
  \bibfield  {author} {\bibinfo {author} {\bibfnamefont {M.~S.}\ \bibnamefont
  {Rudner}}\ and\ \bibinfo {author} {\bibfnamefont {N.~H.}\ \bibnamefont
  {Lindner}},\ }\href {\doibase 10.1038/s42254-020-0170-z} {\bibfield
  {journal} {\bibinfo  {journal} {Nature Reviews Physics}\ }\textbf {\bibinfo
  {volume} {2}},\ \bibinfo {pages} {229} (\bibinfo {year} {2020})}\BibitemShut
  {NoStop}%
\bibitem [{\citenamefont {Morina}\ \emph {et~al.}(2015)\citenamefont {Morina},
  \citenamefont {Kibis}, \citenamefont {Pervishko},\ and\ \citenamefont
  {Shelykh}}]{Shelykh2015a}%
  \BibitemOpen
  \bibfield  {author} {\bibinfo {author} {\bibfnamefont {S.}~\bibnamefont
  {Morina}}, \bibinfo {author} {\bibfnamefont {O.~V.}\ \bibnamefont {Kibis}},
  \bibinfo {author} {\bibfnamefont {A.~A.}\ \bibnamefont {Pervishko}}, \ and\
  \bibinfo {author} {\bibfnamefont {I.~A.}\ \bibnamefont {Shelykh}},\ }\href
  {\doibase 10.1103/PhysRevB.91.155312} {\bibfield  {journal} {\bibinfo
  {journal} {Phys. Rev. B}\ }\textbf {\bibinfo {volume} {91}},\ \bibinfo
  {pages} {155312} (\bibinfo {year} {2015})}\BibitemShut {NoStop}%
\bibitem [{\citenamefont {Pervishko}\ \emph {et~al.}(2015)\citenamefont
  {Pervishko}, \citenamefont {Kibis}, \citenamefont {Morina},\ and\
  \citenamefont {Shelykh}}]{Shelykh2015b}%
  \BibitemOpen
  \bibfield  {author} {\bibinfo {author} {\bibfnamefont {A.~A.}\ \bibnamefont
  {Pervishko}}, \bibinfo {author} {\bibfnamefont {O.~V.}\ \bibnamefont
  {Kibis}}, \bibinfo {author} {\bibfnamefont {S.}~\bibnamefont {Morina}}, \
  and\ \bibinfo {author} {\bibfnamefont {I.~A.}\ \bibnamefont {Shelykh}},\
  }\href {\doibase 10.1103/PhysRevB.92.205403} {\bibfield  {journal} {\bibinfo
  {journal} {Phys. Rev. B}\ }\textbf {\bibinfo {volume} {92}},\ \bibinfo
  {pages} {205403} (\bibinfo {year} {2015})}\BibitemShut {NoStop}%
\bibitem [{\citenamefont {Dini}\ \emph {et~al.}(2016)\citenamefont {Dini},
  \citenamefont {Kibis},\ and\ \citenamefont {Shelykh}}]{Shelykh2016a}%
  \BibitemOpen
  \bibfield  {author} {\bibinfo {author} {\bibfnamefont {K.}~\bibnamefont
  {Dini}}, \bibinfo {author} {\bibfnamefont {O.~V.}\ \bibnamefont {Kibis}}, \
  and\ \bibinfo {author} {\bibfnamefont {I.~A.}\ \bibnamefont {Shelykh}},\
  }\href {\doibase 10.1103/PhysRevB.93.235411} {\bibfield  {journal} {\bibinfo
  {journal} {Phys. Rev. B}\ }\textbf {\bibinfo {volume} {93}},\ \bibinfo
  {pages} {235411} (\bibinfo {year} {2016})}\BibitemShut {NoStop}%
\bibitem [{\citenamefont {Kibis}\ \emph {et~al.}(2016)\citenamefont {Kibis},
  \citenamefont {Morina}, \citenamefont {Dini},\ and\ \citenamefont
  {Shelykh}}]{Shelykh2016b}%
  \BibitemOpen
  \bibfield  {author} {\bibinfo {author} {\bibfnamefont {O.~V.}\ \bibnamefont
  {Kibis}}, \bibinfo {author} {\bibfnamefont {S.}~\bibnamefont {Morina}},
  \bibinfo {author} {\bibfnamefont {K.}~\bibnamefont {Dini}}, \ and\ \bibinfo
  {author} {\bibfnamefont {I.~A.}\ \bibnamefont {Shelykh}},\ }\href {\doibase
  10.1103/PhysRevB.93.115420} {\bibfield  {journal} {\bibinfo  {journal} {Phys.
  Rev. B}\ }\textbf {\bibinfo {volume} {93}},\ \bibinfo {pages} {115420}
  (\bibinfo {year} {2016})}\BibitemShut {NoStop}%
\bibitem [{\citenamefont {Oka}\ and\ \citenamefont {Aoki}(2009)}]{Oka2009}%
  \BibitemOpen
  \bibfield  {author} {\bibinfo {author} {\bibfnamefont {T.}~\bibnamefont
  {Oka}}\ and\ \bibinfo {author} {\bibfnamefont {H.}~\bibnamefont {Aoki}},\
  }\href {\doibase 10.1103/PhysRevB.79.081406} {\bibfield  {journal} {\bibinfo
  {journal} {Phys. Rev. B}\ }\textbf {\bibinfo {volume} {79}},\ \bibinfo
  {pages} {081406(R)} (\bibinfo {year} {2009})}\BibitemShut {NoStop}%
\bibitem [{\citenamefont {Milicevic}\ \emph
  {et~al.}(2017)\citenamefont {Milicevic}, \citenamefont {Ozawa},
  \citenamefont {Montambaux}, \citenamefont {Carusotto}, \citenamefont
  {Galopin}, \citenamefont {Lema\^{\i}tre}, \citenamefont {Le~Gratiet},
  \citenamefont {Sagnes}, \citenamefont {Bloch},\ and\ \citenamefont
  {Amo}}]{1610Montambaux}%
  \BibitemOpen
  \bibfield  {author} {\bibinfo {author} {\bibfnamefont {M.}~\bibnamefont
  {Milicevic}}, \bibinfo {author} {\bibfnamefont {T.}~\bibnamefont {Ozawa}},
  \bibinfo {author} {\bibfnamefont {G.}~\bibnamefont {Montambaux}}, \bibinfo
  {author} {\bibfnamefont {I.}~\bibnamefont {Carusotto}}, \bibinfo {author}
  {\bibfnamefont {E.}~\bibnamefont {Galopin}}, \bibinfo {author} {\bibfnamefont
  {A.}~\bibnamefont {Lema\^{\i}tre}}, \bibinfo {author} {\bibfnamefont
  {L.}~\bibnamefont {Le~Gratiet}}, \bibinfo {author} {\bibfnamefont
  {I.}~\bibnamefont {Sagnes}}, \bibinfo {author} {\bibfnamefont
  {J.}~\bibnamefont {Bloch}}, \ and\ \bibinfo {author} {\bibfnamefont
  {A.}~\bibnamefont {Amo}},\ }\href {\doibase 10.1103/PhysRevLett.118.107403}
  {\bibfield  {journal} {\bibinfo  {journal} {Phys. Rev. Lett.}\ }\textbf
  {\bibinfo {volume} {118}},\ \bibinfo {pages} {107403} (\bibinfo {year}
  {2017})}\BibitemShut {NoStop}%
\bibitem [{\citenamefont {Katz}\ \emph {et~al.}(2019)\citenamefont {Katz},
  \citenamefont {Refael},\ and\ \citenamefont {Lindner}}]{1910.13510_Refael}%
  \BibitemOpen
  \bibfield  {author} {\bibinfo {author} {\bibfnamefont {O.}~\bibnamefont
  {Katz}}, \bibinfo {author} {\bibfnamefont {G.}~\bibnamefont {Refael}}, \ and\
  \bibinfo {author} {\bibfnamefont {N.~H.}\ \bibnamefont {Lindner}},\
  }\href@noop {} {\enquote {\bibinfo {title} {Floquet flat-band engineering of
  twisted bilayer graphene},}\ } (\bibinfo {year} {2019}),\ \Eprint
  {http://arxiv.org/abs/arXiv:1910.13510} {arXiv:1910.13510} \BibitemShut
  {NoStop}%
\bibitem [{\citenamefont {Dal~Lago}\ \emph {et~al.}(2017)\citenamefont
  {Dal~Lago}, \citenamefont {Su\'arez~Morell},\ and\ \citenamefont
  {Foa~Torres}}]{FoaTorres1}%
  \BibitemOpen
  \bibfield  {author} {\bibinfo {author} {\bibfnamefont {V.}~\bibnamefont
  {Dal~Lago}}, \bibinfo {author} {\bibfnamefont {E.}~\bibnamefont
  {Su\'arez~Morell}}, \ and\ \bibinfo {author} {\bibfnamefont {L.~E.~F.}\
  \bibnamefont {Foa~Torres}},\ }\href {\doibase 10.1103/PhysRevB.96.235409}
  {\bibfield  {journal} {\bibinfo  {journal} {Phys. Rev. B}\ }\textbf {\bibinfo
  {volume} {96}},\ \bibinfo {pages} {235409} (\bibinfo {year}
  {2017})}\BibitemShut {NoStop}%
\bibitem [{\citenamefont {Perez-Piskunow}\ \emph {et~al.}(2014)\citenamefont
  {Perez-Piskunow}, \citenamefont {Usaj}, \citenamefont {Balseiro},\ and\
  \citenamefont {Foa Torres}}]{PRB2014_PerezPiskunow}%
  \BibitemOpen
  \bibfield  {author} {\bibinfo {author} {\bibfnamefont {P.~M.}\ \bibnamefont
  {Perez-Piskunow}}, \bibinfo {author} {\bibfnamefont {G.}~\bibnamefont
  {Usaj}}, \bibinfo {author} {\bibfnamefont {C.~A.}\ \bibnamefont {Balseiro}},
  \ and\ \bibinfo {author} {\bibfnamefont {L.~E. F.~F.}\ \bibnamefont
  {Foa Torres}},\ }\href {\doibase 10.1103/PhysRevB.89.121401} {\bibfield
  {journal} {\bibinfo  {journal} {Phys. Rev. B}\ }\textbf {\bibinfo {volume}
  {89}},\ \bibinfo {pages} {121401(R)} (\bibinfo {year} {2014})}\BibitemShut
  {NoStop}%
\bibitem [{\citenamefont {Ezawa}(2012{\natexlab{a}})}]{PRL2012Ezawa}%
  \BibitemOpen
  \bibfield  {author} {\bibinfo {author} {\bibfnamefont {M.}~\bibnamefont
  {Ezawa}},\ }\href {\doibase 10.1103/PhysRevLett.109.055502} {\bibfield
  {journal} {\bibinfo  {journal} {Phys. Rev. Lett.}\ }\textbf {\bibinfo
  {volume} {109}},\ \bibinfo {pages} {055502} (\bibinfo {year}
  {2012}{\natexlab{a}})}\BibitemShut {NoStop}%
\bibitem [{\citenamefont {Ezawa}(2012{\natexlab{b}})}]{NJP2012Ezawa}%
  \BibitemOpen
  \bibfield  {author} {\bibinfo {author} {\bibfnamefont {M.}~\bibnamefont
  {Ezawa}},\ }\href {\doibase 10.1088/1367-2630/14/3/033003} {\bibfield
  {journal} {\bibinfo  {journal} {New Journal of Physics}\ }\textbf {\bibinfo
  {volume} {14}},\ \bibinfo {pages} {033003} (\bibinfo {year}
  {2012}{\natexlab{b}})}\BibitemShut {NoStop}%
\bibitem [{\citenamefont {Ezawa}(2013)}]{PRL2013Ezawa}%
  \BibitemOpen
  \bibfield  {author} {\bibinfo {author} {\bibfnamefont {M.}~\bibnamefont
  {Ezawa}},\ }\href {\doibase 10.1103/PhysRevLett.110.026603} {\bibfield
  {journal} {\bibinfo  {journal} {Phys. Rev. Lett.}\ }\textbf {\bibinfo
  {volume} {110}},\ \bibinfo {pages} {026603} (\bibinfo {year}
  {2013})}\BibitemShut {NoStop}%
\bibitem [{\citenamefont {David}\ \emph {et~al.}(2019)\citenamefont {David},
  \citenamefont {Rakyta}, \citenamefont {Korm\'anyos},\ and\ \citenamefont
  {Burkard}}]{PRB2019_Burkard}%
  \BibitemOpen
  \bibfield  {author} {\bibinfo {author} {\bibfnamefont {A.}~\bibnamefont
  {David}}, \bibinfo {author} {\bibfnamefont {P.}~\bibnamefont {Rakyta}},
  \bibinfo {author} {\bibfnamefont {A.}~\bibnamefont {Korm\'anyos}}, \ and\
  \bibinfo {author} {\bibfnamefont {G.}~\bibnamefont {Burkard}},\ }\href
  {\doibase 10.1103/PhysRevB.100.085412} {\bibfield  {journal} {\bibinfo
  {journal} {Phys. Rev. B}\ }\textbf {\bibinfo {volume} {100}},\ \bibinfo
  {pages} {085412} (\bibinfo {year} {2019})}\BibitemShut {NoStop}%
\bibitem [{\citenamefont {Wang}\ \emph {et~al.}(2013)\citenamefont {Wang},
  \citenamefont {Steinberg}, \citenamefont {Jarillo-Herrero},\ and\
  \citenamefont {Gedik}}]{Wang2013}%
  \BibitemOpen
  \bibfield  {author} {\bibinfo {author} {\bibfnamefont {Y.~H.}\ \bibnamefont
  {Wang}}, \bibinfo {author} {\bibfnamefont {H.}~\bibnamefont {Steinberg}},
  \bibinfo {author} {\bibfnamefont {P.}~\bibnamefont {Jarillo-Herrero}}, \ and\
  \bibinfo {author} {\bibfnamefont {N.}~\bibnamefont {Gedik}},\ }\href
  {\doibase 10.1126/science.1239834} {\bibfield  {journal} {\bibinfo  {journal}
  {Science}\ }\textbf {\bibinfo {volume} {342}},\ \bibinfo {pages} {453}
  (\bibinfo {year} {2013})}\BibitemShut {NoStop}%
\bibitem [{\citenamefont {Calvo}\ \emph {et~al.}(2015)\citenamefont {Calvo},
  \citenamefont {Foa~Torres}, \citenamefont {Perez-Piskunow}, \citenamefont
  {Balseiro},\ and\ \citenamefont {Usaj}}]{Calvo2015}%
  \BibitemOpen
  \bibfield  {author} {\bibinfo {author} {\bibfnamefont {H.~L.}\ \bibnamefont
  {Calvo}}, \bibinfo {author} {\bibfnamefont {L.~E.~F.}\ \bibnamefont
  {Foa~Torres}}, \bibinfo {author} {\bibfnamefont {P.~M.}\ \bibnamefont
  {Perez-Piskunow}}, \bibinfo {author} {\bibfnamefont {C.~A.}\ \bibnamefont
  {Balseiro}}, \ and\ \bibinfo {author} {\bibfnamefont {G.}~\bibnamefont
  {Usaj}},\ }\href {\doibase 10.1103/PhysRevB.91.241404} {\bibfield  {journal}
  {\bibinfo  {journal} {Phys. Rev. B}\ }\textbf {\bibinfo {volume} {91}},\
  \bibinfo {pages} {241404(R)} (\bibinfo {year} {2015})}\BibitemShut {NoStop}%
\bibitem [{\citenamefont {Mahmood}\ \emph {et~al.}(2016)\citenamefont
  {Mahmood}, \citenamefont {Chan}, \citenamefont {Alpichshev}, \citenamefont
  {Gardner}, \citenamefont {Lee}, \citenamefont {Lee},\ and\ \citenamefont
  {Gedik}}]{Mahmood2016}%
  \BibitemOpen
  \bibfield  {author} {\bibinfo {author} {\bibfnamefont {F.}~\bibnamefont
  {Mahmood}}, \bibinfo {author} {\bibfnamefont {C.-K.}\ \bibnamefont {Chan}},
  \bibinfo {author} {\bibfnamefont {Z.}~\bibnamefont {Alpichshev}}, \bibinfo
  {author} {\bibfnamefont {D.}~\bibnamefont {Gardner}}, \bibinfo {author}
  {\bibfnamefont {Y.}~\bibnamefont {Lee}}, \bibinfo {author} {\bibfnamefont
  {P.~A.}\ \bibnamefont {Lee}}, \ and\ \bibinfo {author} {\bibfnamefont
  {N.}~\bibnamefont {Gedik}},\ }\href {\doibase 10.1038/nphys3609} {\bibfield
  {journal} {\bibinfo  {journal} {Nature Physics}\ }\textbf {\bibinfo {volume}
  {12}},\ \bibinfo {pages} {306} (\bibinfo {year} {2016})}\BibitemShut
  {NoStop}%
\bibitem [{\citenamefont {Prange}\ and\ \citenamefont
  {Girvin}(1987)}]{Prange1987}%
  \BibitemOpen
  \bibfield  {author} {\bibinfo {author} {\bibfnamefont {R.~E.}\ \bibnamefont
  {Prange}}\ and\ \bibinfo {author} {\bibfnamefont {S.~M.}\ \bibnamefont
  {Girvin}},\ }\href {\doibase 10.1007/978-1-4612-3350-3} {\emph {\bibinfo
  {title} {The Quantum Hall Effect}}}\ (\bibinfo  {publisher} {Springer-Verlag
  New York},\ \bibinfo {year} {1987})\BibitemShut {NoStop}%
\bibitem [{\citenamefont {Klitzing}\ \emph {et~al.}(1980)\citenamefont
  {Klitzing}, \citenamefont {Dorda},\ and\ \citenamefont
  {Pepper}}]{Klitzing1980}%
  \BibitemOpen
  \bibfield  {author} {\bibinfo {author} {\bibfnamefont {K.~v.}\ \bibnamefont
  {Klitzing}}, \bibinfo {author} {\bibfnamefont {G.}~\bibnamefont {Dorda}}, \
  and\ \bibinfo {author} {\bibfnamefont {M.}~\bibnamefont {Pepper}},\ }\href
  {\doibase 10.1103/PhysRevLett.45.494} {\bibfield  {journal} {\bibinfo
  {journal} {Phys. Rev. Lett.}\ }\textbf {\bibinfo {volume} {45}},\ \bibinfo
  {pages} {494} (\bibinfo {year} {1980})}\BibitemShut {NoStop}%
\bibitem [{\citenamefont {Morimoto}\ \emph {et~al.}(2009)\citenamefont
  {Morimoto}, \citenamefont {Hatsugai},\ and\ \citenamefont
  {Aoki}}]{Morimoto2009}%
  \BibitemOpen
  \bibfield  {author} {\bibinfo {author} {\bibfnamefont {T.}~\bibnamefont
  {Morimoto}}, \bibinfo {author} {\bibfnamefont {Y.}~\bibnamefont {Hatsugai}},
  \ and\ \bibinfo {author} {\bibfnamefont {H.}~\bibnamefont {Aoki}},\ }\href
  {\doibase 10.1103/PhysRevLett.103.116803} {\bibfield  {journal} {\bibinfo
  {journal} {Phys. Rev. Lett.}\ }\textbf {\bibinfo {volume} {103}},\ \bibinfo
  {pages} {116803} (\bibinfo {year} {2009})}\BibitemShut {NoStop}%
\bibitem [{\citenamefont {Hernang\'omez-P\'erez}\ \emph
  {et~al.}(2014)\citenamefont {Hernang\'omez-P\'erez}, \citenamefont
  {Florens},\ and\ \citenamefont {Champel}}]{Hernangomez2014}%
  \BibitemOpen
  \bibfield  {author} {\bibinfo {author} {\bibfnamefont {D.}~\bibnamefont
  {Hernang\'omez-P\'erez}}, \bibinfo {author} {\bibfnamefont {S.}~\bibnamefont
  {Florens}}, \ and\ \bibinfo {author} {\bibfnamefont {T.}~\bibnamefont
  {Champel}},\ }\href {\doibase 10.1103/PhysRevB.89.155314} {\bibfield
  {journal} {\bibinfo  {journal} {Phys. Rev. B}\ }\textbf {\bibinfo {volume}
  {89}},\ \bibinfo {pages} {155314} (\bibinfo {year} {2014})}\BibitemShut
  {NoStop}%
\bibitem [{\citenamefont {Li}\ and\ \citenamefont
  {Carbotte}(2014)}]{Carbotte2014}%
  \BibitemOpen
  \bibfield  {author} {\bibinfo {author} {\bibfnamefont {Z.}~\bibnamefont
  {Li}}\ and\ \bibinfo {author} {\bibfnamefont {J.~P.}\ \bibnamefont
  {Carbotte}},\ }\href {\doibase 10.1103/PhysRevB.89.085413} {\bibfield
  {journal} {\bibinfo  {journal} {Phys. Rev. B}\ }\textbf {\bibinfo {volume}
  {89}},\ \bibinfo {pages} {085413} (\bibinfo {year} {2014})}\BibitemShut
  {NoStop}%
\bibitem [{\citenamefont {Nitta}\ \emph {et~al.}(1997)\citenamefont {Nitta},
  \citenamefont {Akazaki}, \citenamefont {Takayanagi},\ and\ \citenamefont
  {Enoki}}]{Nitta1997}%
  \BibitemOpen
  \bibfield  {author} {\bibinfo {author} {\bibfnamefont {J.}~\bibnamefont
  {Nitta}}, \bibinfo {author} {\bibfnamefont {T.}~\bibnamefont {Akazaki}},
  \bibinfo {author} {\bibfnamefont {H.}~\bibnamefont {Takayanagi}}, \ and\
  \bibinfo {author} {\bibfnamefont {T.}~\bibnamefont {Enoki}},\ }\href
  {\doibase 10.1103/PhysRevLett.78.1335} {\bibfield  {journal} {\bibinfo
  {journal} {Phys. Rev. Lett.}\ }\textbf {\bibinfo {volume} {78}},\ \bibinfo
  {pages} {1335} (\bibinfo {year} {1997})}\BibitemShut {NoStop}%
\bibitem [{\citenamefont {Becker}\ \emph {et~al.}(2010)\citenamefont {Becker},
  \citenamefont {Liebmann}, \citenamefont {Mashoff}, \citenamefont {Pratzer},\
  and\ \citenamefont {Morgenstern}}]{Morgenstern2011}%
  \BibitemOpen
  \bibfield  {author} {\bibinfo {author} {\bibfnamefont {S.}~\bibnamefont
  {Becker}}, \bibinfo {author} {\bibfnamefont {M.}~\bibnamefont {Liebmann}},
  \bibinfo {author} {\bibfnamefont {T.}~\bibnamefont {Mashoff}}, \bibinfo
  {author} {\bibfnamefont {M.}~\bibnamefont {Pratzer}}, \ and\ \bibinfo
  {author} {\bibfnamefont {M.}~\bibnamefont {Morgenstern}},\ }\href {\doibase
  10.1103/PhysRevB.81.155308} {\bibfield  {journal} {\bibinfo  {journal} {Phys.
  Rev. B}\ }\textbf {\bibinfo {volume} {81}},\ \bibinfo {pages} {155308}
  (\bibinfo {year} {2010})}\BibitemShut {NoStop}%
\bibitem [{\citenamefont {Singh}\ and\ \citenamefont
  {Romero}(2017)}]{PRB2017Singh}%
  \BibitemOpen
  \bibfield  {author} {\bibinfo {author} {\bibfnamefont {S.}~\bibnamefont
  {Singh}}\ and\ \bibinfo {author} {\bibfnamefont {A.~H.}\ \bibnamefont
  {Romero}},\ }\href {\doibase 10.1103/PhysRevB.95.165444} {\bibfield
  {journal} {\bibinfo  {journal} {Phys. Rev. B}\ }\textbf {\bibinfo {volume}
  {95}},\ \bibinfo {pages} {165444} (\bibinfo {year} {2017})}\BibitemShut
  {NoStop}%
\bibitem [{\citenamefont {Debald}\ and\ \citenamefont
  {Emary}(2005)}]{Debald2005}%
  \BibitemOpen
  \bibfield  {author} {\bibinfo {author} {\bibfnamefont {S.}~\bibnamefont
  {Debald}}\ and\ \bibinfo {author} {\bibfnamefont {C.}~\bibnamefont {Emary}},\
  }\href {\doibase 10.1103/PhysRevLett.94.226803} {\bibfield  {journal}
  {\bibinfo  {journal} {Phys. Rev. Lett.}\ }\textbf {\bibinfo {volume} {94}},\
  \bibinfo {pages} {226803} (\bibinfo {year} {2005})}\BibitemShut {NoStop}%
\bibitem [{\citenamefont {Bychkov}\ and\ \citenamefont
  {Rashba}(1984)}]{Bychkov1984}%
  \BibitemOpen
  \bibfield  {author} {\bibinfo {author} {\bibfnamefont {Y.~A.}\ \bibnamefont
  {Bychkov}}\ and\ \bibinfo {author} {\bibfnamefont {E.~I.}\ \bibnamefont
  {Rashba}},\ }\href {\doibase 10.1088/0022-3719/17/33/015} {\bibfield
  {journal} {\bibinfo  {journal} {Journal of Physics C: Solid State Physics}\
  }\textbf {\bibinfo {volume} {17}},\ \bibinfo {pages} {6039} (\bibinfo {year}
  {1984})}\BibitemShut {NoStop}%
\bibitem [{\citenamefont {Hernang\'omez-P\'erez}\ \emph
  {et~al.}(2013)\citenamefont {Hernang\'omez-P\'erez}, \citenamefont {Ulrich},
  \citenamefont {Florens},\ and\ \citenamefont {Champel}}]{Hernangomez2013}%
  \BibitemOpen
  \bibfield  {author} {\bibinfo {author} {\bibfnamefont {D.}~\bibnamefont
  {Hernang\'omez-P\'erez}}, \bibinfo {author} {\bibfnamefont {J.}~\bibnamefont
  {Ulrich}}, \bibinfo {author} {\bibfnamefont {S.}~\bibnamefont {Florens}}, \
  and\ \bibinfo {author} {\bibfnamefont {T.}~\bibnamefont {Champel}},\ }\href
  {\doibase 10.1103/PhysRevB.88.245433} {\bibfield  {journal} {\bibinfo
  {journal} {Phys. Rev. B}\ }\textbf {\bibinfo {volume} {88}},\ \bibinfo
  {pages} {245433} (\bibinfo {year} {2013})}\BibitemShut {NoStop}%
\bibitem [{\citenamefont {Champel}\ and\ \citenamefont
  {Florens}(2010)}]{Champel2010}%
  \BibitemOpen
  \bibfield  {author} {\bibinfo {author} {\bibfnamefont {T.}~\bibnamefont
  {Champel}}\ and\ \bibinfo {author} {\bibfnamefont {S.}~\bibnamefont
  {Florens}},\ }\href {\doibase 10.1103/PhysRevB.82.045421} {\bibfield
  {journal} {\bibinfo  {journal} {Phys. Rev. B}\ }\textbf {\bibinfo {volume}
  {82}},\ \bibinfo {pages} {045421} (\bibinfo {year} {2010})}\BibitemShut
  {NoStop}%
\bibitem [{\citenamefont {L\'opez}\ \emph {et~al.}(2015)\citenamefont
  {L\'opez}, \citenamefont {Di~Teodoro}, \citenamefont {Schliemann},
  \citenamefont {Berche},\ and\ \citenamefont {Santos}}]{Lopez2015}%
  \BibitemOpen
  \bibfield  {author} {\bibinfo {author} {\bibfnamefont {A.}~\bibnamefont
  {L\'opez}}, \bibinfo {author} {\bibfnamefont {A.}~\bibnamefont {Di~Teodoro}},
  \bibinfo {author} {\bibfnamefont {J.}~\bibnamefont {Schliemann}}, \bibinfo
  {author} {\bibfnamefont {B.}~\bibnamefont {Berche}}, \ and\ \bibinfo {author}
  {\bibfnamefont {B.}~\bibnamefont {Santos}},\ }\href {\doibase
  10.1103/PhysRevB.92.235411} {\bibfield  {journal} {\bibinfo  {journal} {Phys.
  Rev. B}\ }\textbf {\bibinfo {volume} {92}},\ \bibinfo {pages} {235411}
  (\bibinfo {year} {2015})}\BibitemShut {NoStop}%
\bibitem [{\citenamefont {Zhou}\ and\ \citenamefont {Wu}(2011)}]{Wu2011}%
  \BibitemOpen
  \bibfield  {author} {\bibinfo {author} {\bibfnamefont {Y.}~\bibnamefont
  {Zhou}}\ and\ \bibinfo {author} {\bibfnamefont {M.~W.}\ \bibnamefont {Wu}},\
  }\href {\doibase 10.1103/PhysRevB.83.245436} {\bibfield  {journal} {\bibinfo
  {journal} {Phys. Rev. B}\ }\textbf {\bibinfo {volume} {83}},\ \bibinfo
  {pages} {245436} (\bibinfo {year} {2011})}\BibitemShut {NoStop}%
\bibitem [{\citenamefont {Scholz}\ \emph {et~al.}(2013)\citenamefont {Scholz},
  \citenamefont {L\'opez},\ and\ \citenamefont {Schliemann}}]{Scholz2013}%
  \BibitemOpen
  \bibfield  {author} {\bibinfo {author} {\bibfnamefont {A.}~\bibnamefont
  {Scholz}}, \bibinfo {author} {\bibfnamefont {A.}~\bibnamefont {L\'opez}}, \
  and\ \bibinfo {author} {\bibfnamefont {J.}~\bibnamefont {Schliemann}},\
  }\href {\doibase 10.1103/PhysRevB.88.045118} {\bibfield  {journal} {\bibinfo
  {journal} {Phys. Rev. B}\ }\textbf {\bibinfo {volume} {88}},\ \bibinfo
  {pages} {045118} (\bibinfo {year} {2013})}\BibitemShut {NoStop}%
\bibitem [{\citenamefont {Grifoni}\ and\ \citenamefont
  {H\"anggi}(1998)}]{Grifoni2009}%
  \BibitemOpen
  \bibfield  {author} {\bibinfo {author} {\bibfnamefont {M.}~\bibnamefont
  {Grifoni}}\ and\ \bibinfo {author} {\bibfnamefont {P.}~\bibnamefont
  {H\"anggi}},\ }\href {\doibase https://doi.org/10.1016/S0370-1573(98)00022-2}
  {\bibfield  {journal} {\bibinfo  {journal} {Physics Reports}\ }\textbf
  {\bibinfo {volume} {304}},\ \bibinfo {pages} {229 } (\bibinfo {year}
  {1998})}\BibitemShut {NoStop}%
\bibitem [{\citenamefont {Karch}\ \emph {et~al.}(2010)\citenamefont {Karch},
  \citenamefont {Olbrich}, \citenamefont {Schmalzbauer}, \citenamefont {Zoth},
  \citenamefont {Brinsteiner}, \citenamefont {Fehrenbacher}, \citenamefont
  {Wurstbauer}, \citenamefont {Glazov}, \citenamefont {Tarasenko},
  \citenamefont {Ivchenko}, \citenamefont {Weiss}, \citenamefont {Eroms},
  \citenamefont {Yakimova}, \citenamefont {Lara-Avila}, \citenamefont
  {Kubatkin},\ and\ \citenamefont {Ganichev}}]{PRLKarch2010}%
  \BibitemOpen
  \bibfield  {author} {\bibinfo {author} {\bibfnamefont {J.}~\bibnamefont
  {Karch}}, \bibinfo {author} {\bibfnamefont {P.}~\bibnamefont {Olbrich}},
  \bibinfo {author} {\bibfnamefont {M.}~\bibnamefont {Schmalzbauer}}, \bibinfo
  {author} {\bibfnamefont {C.}~\bibnamefont {Zoth}}, \bibinfo {author}
  {\bibfnamefont {C.}~\bibnamefont {Brinsteiner}}, \bibinfo {author}
  {\bibfnamefont {M.}~\bibnamefont {Fehrenbacher}}, \bibinfo {author}
  {\bibfnamefont {U.}~\bibnamefont {Wurstbauer}}, \bibinfo {author}
  {\bibfnamefont {M.~M.}\ \bibnamefont {Glazov}}, \bibinfo {author}
  {\bibfnamefont {S.~A.}\ \bibnamefont {Tarasenko}}, \bibinfo {author}
  {\bibfnamefont {E.~L.}\ \bibnamefont {Ivchenko}}, \bibinfo {author}
  {\bibfnamefont {D.}~\bibnamefont {Weiss}}, \bibinfo {author} {\bibfnamefont
  {J.}~\bibnamefont {Eroms}}, \bibinfo {author} {\bibfnamefont
  {R.}~\bibnamefont {Yakimova}}, \bibinfo {author} {\bibfnamefont
  {S.}~\bibnamefont {Lara-Avila}}, \bibinfo {author} {\bibfnamefont
  {S.}~\bibnamefont {Kubatkin}}, \ and\ \bibinfo {author} {\bibfnamefont
  {S.~D.}\ \bibnamefont {Ganichev}},\ }\href {\doibase
  10.1103/PhysRevLett.105.227402} {\bibfield  {journal} {\bibinfo  {journal}
  {Phys. Rev. Lett.}\ }\textbf {\bibinfo {volume} {105}},\ \bibinfo {pages}
  {227402} (\bibinfo {year} {2010})}\BibitemShut {NoStop}%
\bibitem [{\citenamefont {Agarwal}(2012)}]{agarwal_2012}%
  \BibitemOpen
  \bibfield  {author} {\bibinfo {author} {\bibfnamefont {G.~S.}\ \bibnamefont
  {Agarwal}},\ }\href {\doibase 10.1017/CBO9781139035170} {\emph {\bibinfo
  {title} {Quantum Optics}}}\ (\bibinfo  {publisher} {Cambridge University
  Press},\ \bibinfo {year} {2012})\BibitemShut {NoStop}%
\bibitem [{\citenamefont {Romera}\ and\ \citenamefont {de~los
  Santos}(2009)}]{Romera2009}%
  \BibitemOpen
  \bibfield  {author} {\bibinfo {author} {\bibfnamefont {E.}~\bibnamefont
  {Romera}}\ and\ \bibinfo {author} {\bibfnamefont {F.}~\bibnamefont {de~los
  Santos}},\ }\href {\doibase 10.1103/PhysRevB.80.165416} {\bibfield  {journal}
  {\bibinfo  {journal} {Phys. Rev. B}\ }\textbf {\bibinfo {volume} {80}},\
  \bibinfo {pages} {165416} (\bibinfo {year} {2009})}\BibitemShut {NoStop}%
\bibitem [{\citenamefont {Romera}(2011)}]{Romera2011}%
  \BibitemOpen
  \bibfield  {author} {\bibinfo {author} {\bibfnamefont {E.}~\bibnamefont
  {Romera}},\ }\href {\doibase 10.1103/PhysRevA.84.052102} {\bibfield
  {journal} {\bibinfo  {journal} {Phys. Rev. A}\ }\textbf {\bibinfo {volume}
  {84}},\ \bibinfo {pages} {052102} (\bibinfo {year} {2011})}\BibitemShut
  {NoStop}%
\bibitem [{\citenamefont {Torres}\ and\ \citenamefont
  {Kunold}(2005)}]{Kunold2005}%
  \BibitemOpen
  \bibfield  {author} {\bibinfo {author} {\bibfnamefont {M.}~\bibnamefont
  {Torres}}\ and\ \bibinfo {author} {\bibfnamefont {A.}~\bibnamefont
  {Kunold}},\ }\href {\doibase 10.1103/PhysRevB.71.115313} {\bibfield
  {journal} {\bibinfo  {journal} {Phys. Rev. B}\ }\textbf {\bibinfo {volume}
  {71}},\ \bibinfo {pages} {115313} (\bibinfo {year} {2005})}\BibitemShut
  {NoStop}%
\bibitem [{\citenamefont {Li}\ and\ \citenamefont
  {Carbotte}(2013)}]{Carbotte2013}%
  \BibitemOpen
  \bibfield  {author} {\bibinfo {author} {\bibfnamefont {Z.}~\bibnamefont
  {Li}}\ and\ \bibinfo {author} {\bibfnamefont {J.~P.}\ \bibnamefont
  {Carbotte}},\ }\href {\doibase 10.1103/PhysRevB.88.045414} {\bibfield
  {journal} {\bibinfo  {journal} {Phys. Rev. B}\ }\textbf {\bibinfo {volume}
  {88}},\ \bibinfo {pages} {045414} (\bibinfo {year} {2013})}\BibitemShut
  {NoStop}%
\bibitem [{\citenamefont {Dehghani}\ and\ \citenamefont
  {Mitra}(2015)}]{Mitra2015}%
  \BibitemOpen
  \bibfield  {author} {\bibinfo {author} {\bibfnamefont {H.}~\bibnamefont
  {Dehghani}}\ and\ \bibinfo {author} {\bibfnamefont {A.}~\bibnamefont
  {Mitra}},\ }\href {\doibase 10.1103/PhysRevB.92.165111} {\bibfield  {journal}
  {\bibinfo  {journal} {Phys. Rev. B}\ }\textbf {\bibinfo {volume} {92}},\
  \bibinfo {pages} {165111} (\bibinfo {year} {2015})}\BibitemShut {NoStop}%
\bibitem [{\citenamefont {Zubair}\ \emph {et~al.}(2018)\citenamefont {Zubair},
  \citenamefont {Tahir},\ and\ \citenamefont {Vasilopoulos}}]{Zubair2018}%
  \BibitemOpen
  \bibfield  {author} {\bibinfo {author} {\bibfnamefont {M.}~\bibnamefont
  {Zubair}}, \bibinfo {author} {\bibfnamefont {M.}~\bibnamefont {Tahir}}, \
  and\ \bibinfo {author} {\bibfnamefont {P.}~\bibnamefont {Vasilopoulos}},\
  }\href {\doibase 10.1103/PhysRevB.98.155402} {\bibfield  {journal} {\bibinfo
  {journal} {Phys. Rev. B}\ }\textbf {\bibinfo {volume} {98}},\ \bibinfo
  {pages} {155402} (\bibinfo {year} {2018})}\BibitemShut {NoStop}%
\bibitem [{\citenamefont {Dehghani}\ \emph {et~al.}(2015)\citenamefont
  {Dehghani}, \citenamefont {Oka},\ and\ \citenamefont {Mitra}}]{Oka2015}%
  \BibitemOpen
  \bibfield  {author} {\bibinfo {author} {\bibfnamefont {H.}~\bibnamefont
  {Dehghani}}, \bibinfo {author} {\bibfnamefont {T.}~\bibnamefont {Oka}}, \
  and\ \bibinfo {author} {\bibfnamefont {A.}~\bibnamefont {Mitra}},\ }\href
  {\doibase 10.1103/PhysRevB.91.155422} {\bibfield  {journal} {\bibinfo
  {journal} {Phys. Rev. B}\ }\textbf {\bibinfo {volume} {91}},\ \bibinfo
  {pages} {155422} (\bibinfo {year} {2015})}\BibitemShut {NoStop}%
\bibitem [{\citenamefont {Peralta~Gavensky}\ \emph {et~al.}(2018)\citenamefont
  {Peralta~Gavensky}, \citenamefont {Usaj},\ and\ \citenamefont
  {Balseiro}}]{Balseiro2018}%
  \BibitemOpen
  \bibfield  {author} {\bibinfo {author} {\bibfnamefont {L.}~\bibnamefont
  {Peralta~Gavensky}}, \bibinfo {author} {\bibfnamefont {G.}~\bibnamefont
  {Usaj}}, \ and\ \bibinfo {author} {\bibfnamefont {C.~A.}\ \bibnamefont
  {Balseiro}},\ }\href {\doibase 10.1103/PhysRevB.98.165414} {\bibfield
  {journal} {\bibinfo  {journal} {Phys. Rev. B}\ }\textbf {\bibinfo {volume}
  {98}},\ \bibinfo {pages} {165414} (\bibinfo {year} {2018})}\BibitemShut
  {NoStop}%
\bibitem [{Note1()}]{Note1}%
  \BibitemOpen
  \bibinfo {note} {For simplicity, we assume a vanishing Zeeman field, $\Delta
  = 0$, taking into consideration a non-vanishing Zeeman field being
  straightforward.}\BibitemShut {Stop}%
\bibitem [{\citenamefont {Gusynin}\ and\ \citenamefont
  {Sharapov}(2005)}]{Gusynin2005}%
  \BibitemOpen
  \bibfield  {author} {\bibinfo {author} {\bibfnamefont {V.~P.}\ \bibnamefont
  {Gusynin}}\ and\ \bibinfo {author} {\bibfnamefont {S.~G.}\ \bibnamefont
  {Sharapov}},\ }\href {\doibase 10.1103/PhysRevLett.95.146801} {\bibfield
  {journal} {\bibinfo  {journal} {Phys. Rev. Lett.}\ }\textbf {\bibinfo
  {volume} {95}},\ \bibinfo {pages} {146801} (\bibinfo {year}
  {2005})}\BibitemShut {NoStop}%
\bibitem [{\citenamefont {Zyuzin}\ and\ \citenamefont
  {Burkov}(2011)}]{Burkov2011}%
  \BibitemOpen
  \bibfield  {author} {\bibinfo {author} {\bibfnamefont {A.~A.}\ \bibnamefont
  {Zyuzin}}\ and\ \bibinfo {author} {\bibfnamefont {A.~A.}\ \bibnamefont
  {Burkov}},\ }\href {\doibase 10.1103/PhysRevB.83.195413} {\bibfield
  {journal} {\bibinfo  {journal} {Phys. Rev. B}\ }\textbf {\bibinfo {volume}
  {83}},\ \bibinfo {pages} {195413} (\bibinfo {year} {2011})}\BibitemShut
  {NoStop}%
\bibitem [{\citenamefont {K\"onig}\ \emph {et~al.}(2014)\citenamefont
  {K\"onig}, \citenamefont {Ostrovsky}, \citenamefont {Protopopov},
  \citenamefont {Gornyi}, \citenamefont {Burmistrov},\ and\ \citenamefont
  {Mirlin}}]{Konig2014}%
  \BibitemOpen
  \bibfield  {author} {\bibinfo {author} {\bibfnamefont {E.~J.}\ \bibnamefont
  {K\"onig}}, \bibinfo {author} {\bibfnamefont {P.~M.}\ \bibnamefont
  {Ostrovsky}}, \bibinfo {author} {\bibfnamefont {I.~V.}\ \bibnamefont
  {Protopopov}}, \bibinfo {author} {\bibfnamefont {I.~V.}\ \bibnamefont
  {Gornyi}}, \bibinfo {author} {\bibfnamefont {I.~S.}\ \bibnamefont
  {Burmistrov}}, \ and\ \bibinfo {author} {\bibfnamefont {A.~D.}\ \bibnamefont
  {Mirlin}},\ }\href {\doibase 10.1103/PhysRevB.90.165435} {\bibfield
  {journal} {\bibinfo  {journal} {Phys. Rev. B}\ }\textbf {\bibinfo {volume}
  {90}},\ \bibinfo {pages} {165435} (\bibinfo {year} {2014})}\BibitemShut
  {NoStop}%
\bibitem [{\citenamefont {Yudin}\ and\ \citenamefont
  {Shelykh}(2016)}]{Yudin2016}%
  \BibitemOpen
  \bibfield  {author} {\bibinfo {author} {\bibfnamefont {D.}~\bibnamefont
  {Yudin}}\ and\ \bibinfo {author} {\bibfnamefont {I.~A.}\ \bibnamefont
  {Shelykh}},\ }\href {\doibase 10.1103/PhysRevB.94.161404} {\bibfield
  {journal} {\bibinfo  {journal} {Phys. Rev. B}\ }\textbf {\bibinfo {volume}
  {94}},\ \bibinfo {pages} {161404(R)} (\bibinfo {year} {2016})}\BibitemShut
  {NoStop}%
\bibitem [{\citenamefont {Usachov}\ \emph {et~al.}(2015)\citenamefont
  {Usachov}, \citenamefont {Fedorov}, \citenamefont {Otrokov}, \citenamefont
  {Chikina}, \citenamefont {Vilkov}, \citenamefont {Petukhov}, \citenamefont
  {Rybkin}, \citenamefont {Koroteev}, \citenamefont {Chulkov}, \citenamefont
  {Adamchuk}, \citenamefont {Grüneis}, \citenamefont {Laubschat},\ and\
  \citenamefont {Vyalikh}}]{doi:10.1021/nl504693u}%
  \BibitemOpen
  \bibfield  {author} {\bibinfo {author} {\bibfnamefont {D.}~\bibnamefont
  {Usachov}}, \bibinfo {author} {\bibfnamefont {A.}~\bibnamefont {Fedorov}},
  \bibinfo {author} {\bibfnamefont {M.~M.}\ \bibnamefont {Otrokov}}, \bibinfo
  {author} {\bibfnamefont {A.}~\bibnamefont {Chikina}}, \bibinfo {author}
  {\bibfnamefont {O.}~\bibnamefont {Vilkov}}, \bibinfo {author} {\bibfnamefont
  {A.}~\bibnamefont {Petukhov}}, \bibinfo {author} {\bibfnamefont {A.~G.}\
  \bibnamefont {Rybkin}}, \bibinfo {author} {\bibfnamefont {Y.~M.}\
  \bibnamefont {Koroteev}}, \bibinfo {author} {\bibfnamefont {E.~V.}\
  \bibnamefont {Chulkov}}, \bibinfo {author} {\bibfnamefont {V.~K.}\
  \bibnamefont {Adamchuk}}, \bibinfo {author} {\bibfnamefont {A.}~\bibnamefont
  {Grüneis}}, \bibinfo {author} {\bibfnamefont {C.}~\bibnamefont {Laubschat}},
  \ and\ \bibinfo {author} {\bibfnamefont {D.~V.}\ \bibnamefont {Vyalikh}},\
  }\href {\doibase 10.1021/nl504693u} {\bibfield  {journal} {\bibinfo
  {journal} {Nano Letters}\ }\textbf {\bibinfo {volume} {15}},\ \bibinfo
  {pages} {2396} (\bibinfo {year} {2015})},\ \bibinfo {note} {pMID: 25734657},\
  \Eprint {http://arxiv.org/abs/https://doi.org/10.1021/nl504693u}
  {https://doi.org/10.1021/nl504693u} \BibitemShut {NoStop}%
\bibitem [{\citenamefont {Bayer}\ \emph {et~al.}(2019)\citenamefont {Bayer},
  \citenamefont {Gräfing}, \citenamefont {Kerbstadt}, \citenamefont {Pengel},
  \citenamefont {Eickhoff}, \citenamefont {Englert},\ and\ \citenamefont
  {Wollenhaupt}}]{Bayer_2019}%
  \BibitemOpen
  \bibfield  {author} {\bibinfo {author} {\bibfnamefont {T.}~\bibnamefont
  {Bayer}}, \bibinfo {author} {\bibfnamefont {D.}~\bibnamefont {Gräfing}},
  \bibinfo {author} {\bibfnamefont {S.}~\bibnamefont {Kerbstadt}}, \bibinfo
  {author} {\bibfnamefont {D.}~\bibnamefont {Pengel}}, \bibinfo {author}
  {\bibfnamefont {K.}~\bibnamefont {Eickhoff}}, \bibinfo {author}
  {\bibfnamefont {L.}~\bibnamefont {Englert}}, \ and\ \bibinfo {author}
  {\bibfnamefont {M.}~\bibnamefont {Wollenhaupt}},\ }\href {\doibase
  10.1088/1367-2630/aafb87} {\bibfield  {journal} {\bibinfo  {journal} {New
  Journal of Physics}\ }\textbf {\bibinfo {volume} {21}},\ \bibinfo {pages}
  {033001} (\bibinfo {year} {2019})}\BibitemShut {NoStop}%
\bibitem [{\citenamefont {Mak}\ \emph {et~al.}(2008)\citenamefont {Mak},
  \citenamefont {Sfeir}, \citenamefont {Wu}, \citenamefont {Lui}, \citenamefont
  {Misewich},\ and\ \citenamefont {Heinz}}]{Mak2008}%
  \BibitemOpen
  \bibfield  {author} {\bibinfo {author} {\bibfnamefont {K.~F.}\ \bibnamefont
  {Mak}}, \bibinfo {author} {\bibfnamefont {M.~Y.}\ \bibnamefont {Sfeir}},
  \bibinfo {author} {\bibfnamefont {Y.}~\bibnamefont {Wu}}, \bibinfo {author}
  {\bibfnamefont {C.~H.}\ \bibnamefont {Lui}}, \bibinfo {author} {\bibfnamefont
  {J.~A.}\ \bibnamefont {Misewich}}, \ and\ \bibinfo {author} {\bibfnamefont
  {T.~F.}\ \bibnamefont {Heinz}},\ }\href {\doibase
  10.1103/PhysRevLett.101.196405} {\bibfield  {journal} {\bibinfo  {journal}
  {Phys. Rev. Lett.}\ }\textbf {\bibinfo {volume} {101}},\ \bibinfo {pages}
  {196405} (\bibinfo {year} {2008})}\BibitemShut {NoStop}%
\end{thebibliography}

\end{document}